\documentclass[fleqn,usenatbib,useAMS]{mnras}

\usepackage{graphicx}	% Including figure files
\usepackage{amsmath}	% Advanced maths commands
\usepackage{multicol}        % Multi-column entries in tables
\usepackage{bm}		% Bold maths symbols, including upright Greek
\usepackage{pdflscape}	% Landscape pages

\usepackage[T1]{fontenc}
\usepackage{ae,aecompl}

\usepackage{newtxtext,newtxmath}
\newcommand{\kepler}{\emph{Kepler}~}
\newcommand{\gaia}{\emph{Gaia}~}

\newcommand{\logg}{\ensuremath{\log g}~}

\newcommand{\dnu}{\ensuremath{\Delta\nu}~}
\newcommand{\numax}{\ensuremath{\nu_\textup{max}}~}

\newcommand{\rgc}{$R_{GC}$~}
\newcommand{\alfa}{$\alpha$}

\newcommand{\quadraaperta}{[}
\newcommand{\quadrachiusa}{]}

\title[\quadraaperta C/N\quadrachiusa~ as a chemical clock]{Stellar ages from [C/N] in giant stars: applicability and limitations}

\author[G. Casali et al.]{
G. Casali$^{1,2}$
\thanks{E-mail: giada.casali@anu.edu.au},
L. Casagrande$^{1}$,
F. Vincenzo$^{3}$,
A. Miglio$^{4}$,
M. Ness$^{1}$,
G. Tautvaišiene$^{5}$,
L. Magrini$^{6}$,
S. Buder$^{1}$
\\
$^{1}$Research School of Astronomy and Astrophysics, The Australian National University, Canberra, ACT 2611, Australia\\
$^{2}$INAF - Osservatorio di Astrofisica e Scienza dello Spazio, Via P. Gobetti 93/3, 40129, Bologna, Italy\\
$^{3}$Dipartimento di Fisica e Astronomia "Ettore Majorana", Università degli Studi di Catania, via S. Sofia 64, 95125, Catania, Italy\\
$^{4}$Dipartimento di Fisica e Astronomia, Università degli Studi di Bologna, Via Gobetti 93/2, I-40129, Bologna, Italy\\
$^{5}$Vilnius University, Faculty of Physics, Institute of Theoretical Physics and Astronomy, Sauletekio av. 3, 10257, Vilnius, Lithuania\\
$^{6}$INAF - Osservatorio Astrofisico di Arcetri, Largo E. Fermi, 3, 50125 Firenze, Italy\\
}
\date{Accepted 2026 September 12. Received 2026 September 09; in original form 2026 July 02
}

\pubyear{{\the\year{}}}

\begin{document}
\label{firstpage}
\pagerange{\pageref{firstpage}--\pageref{lastpage}}
\maketitle


\begin{abstract}
The [C/N] abundance ratio is a powerful age indicator for giant stars, enabling age estimates for spectroscopic samples where traditional methods are limited. We calibrate a multivariate [C/N]-age-[Fe/H] relationship using 9,122 giant stars observed by \emph{Kepler}, with asteroseismic ages and APOGEE DR17 abundances. The calibration is performed separately for lower red giant branch (LRGB), upper RGB (URGB), and red clump (RC) stars. 
We validate the relationships using independent samples from K2, TESS, and open clusters, finding good agreement with reference ages, particularly for LRGB stars, with typical precision of $\sim$30\% for ages between $\sim$2 - 10 Gyr. The performance degrades for URGB and RC stars and low metallicities, likely due to extra mixing processes, for young ($\lesssim$2 Gyr) and old ($\gtrsim$10 Gyr) stars, where the [C/N]-age correlation weakens.
We show that the [C/N]-based ages reproduce known Milky Way features, including the age distribution across the [$\alpha$/Fe]-[Fe/H] plane and the flaring of the Galactic disc. Compared to other chemical clocks, [C/N] provides more robust and precise age estimates within its domain of validity. 
Our results demonstrate that [C/N], combined with metallicity, is an effective empirical tool for deriving stellar ages of giant stars from spectroscopic surveys, particularly for LRGB stars with [Fe/H] $>-1$, enabling Galactic archaeology studies beyond the reach of current asteroseismic samples. %\SB{I think that might be me, but I think lower RGB and upper RGB is maybe easier to digest (and not much longer than LRGB and URGB).}
\end{abstract}

\begin{keywords}
Galaxy: evolution -- Galaxy: abundances -- Galaxy: disc -- stars: abundances -- stars: late-type –asteroseismology
\end{keywords}

%%%%%%%%%%%%%%%%%%%%%%%%%%%%%%%%%%%%%%%%%%%%%%%%%%

%%%%%%%%%%%%%%%%% BODY OF PAPER %%%%%%%%%%%%%%%%%%

% The MNRAS class isn't designed to include a table of contents, but for this document one is useful.
% I therefore have to do some kludging to make it work without masses of blank space.
%\begingroup
%\let\clearpage\relax
%\tableofcontents
%\endgroup
%\newpage

\section{Introduction}
Determining precise and accurate ages for large numbers of stars in the Milky Way is crucial to study the Galaxy formation and evolution and compare observed data with galaxy formation simulations.
However, stellar ages are among the most difficult quantities to determine in astrophysics because they cannot be directly inferred,  unlike, e.g., temperature and distance \citep[see e.g.][]{soderblom10}.

The most used age-determination method is the isochrone fitting: comparing the location of stars in the Hertzsprung-Russell diagram (HRD) with theoretical models. This technique works really well with star clusters, less with field stars. Indeed, it determines precise ages in regions of the HRD where isochrones of different ages are clearly separated (e.g. main-sequence turn-off or subgiant branch). However, stellar ages are very hard to determine in regions where the isochrones of different ages are very close, such as the red giant branch (RGB) and low-main sequence.  
Nevertheless, giant stars are bright and can be observed at great distances from us, allowing to study the structure of the Milky Way. For this reason, giant stars are the main targets of large spectroscopic surveys \citep[e.g.,][]{apogeedr17,randich22}.

Among the alternative methods to estimate ages with respect to the isochrone fitting, one of the most powerful is asteroseismology. Through asteroseismology, we have been able to detect solar-like pulsations in thousands of G-K giants using data collected by the COnvection ROtation and planetary Transits \citep[CoRoT,][]{baglin06}, \kepler \citep{gilliland10}, K2 \citep{howell14}, and Transiting Exoplanet Survey Satellite \citep[TESS,][]{ricker14} missions.
Stellar pulsation frequencies are connected to a star's internal structure, offering constraints on its fundamental properties (radius, mass, and age) and evolutionary state \citep[see][for a review]{chaplin13}. In particular, two global asteroseismic observables -- the average large frequency separation ($\Delta\nu$) and the frequency of maximum oscillation power ($\nu_{\rm max}$) -- can be used to infer stellar parameters. 
%Another method to get the highest levels of precision from asteroseismology is obtained when comparing observed individual mode frequencies (tens of constraints, compared to the two provided by the global seismic parameters) to stellar models \citep[e.g.,][]{huber13,lebreton14,lillobox14,davies16,silvaaguirre17,montalban21}.

However, asteroseismic data are currently available only for relatively small samples of stars, confined to a few fields across the Milky Way.
For this reason, we look forward to  alternative approaches to determining stellar ages to get ages for large samples of stars. One of these approaches relies on the relationship between ages and chemical abundances, which can be applied to hundreds of thousands of stars provided by large spectroscopic surveys. This technique is based on the so-called chemical clocks: abundance ratios between elements that vary in opposite directions with time. These elements are produced on different timescales owing to their distinct nucleosynthetic sites \citep[e.g. Y/Mg;][and references therein]{dasilva12,nissen15,casamiquela21,casali25}, or are modified in opposite ways by stellar evolution \citep[e.g. C/N;][]{masseron15,salaris15}. Their ratios are constructed to maximise the correlation with stellar age.

Among the various chemical clocks, the most effective one for giant stars is the [C/N] ratio.
Carbon and nitrogen are processed through the CNO-cycle in previous evolutionary phases before the RGB and are taken towards the surface by penetration of the convective envelope into the stellar interior \citep{iben65}. The phase in which the convective envelope deepens, reaching into layers where nuclear fusion has already altered the chemical composition, is called the first dredge-up (FDU, hereafter). As a result of this convective mixing, the atmosphere shows a variation in the chemical composition, which in particular changes the abundance ratio [C/N]. Because the penetration of convection in the inner regions, and therefore the abundances of C and N that are brought to the stellar surface, depends on the stellar mass and because the mass is related to the age in the RGB phase, the [C/N] ratio can be used to estimate stellar ages \citep{masseron15,salaris15,ness16,lagarde17,casali19,spoo22,grazina25,spoo25,roberts25,pakstiene26,curjuric26}.

Previous studies have investigated the use of [C/N] ratio as a chemical clock. \citet{martig16} derived stellar ages for about 52,000 stars in APOGEE DR12 by calibrating [C/N] using a training sample of 1,475 red giants with asteroseismic masses and ages from the APOKASC survey \citep[a spectroscopic follow-up by APOGEE of stars with asteroseismic data from the Kepler Asteroseismic Science Consortium,][]{pinsonneault14} with abundances from APOGEE DR12. However, \citet{mackereth17} later showed that the \citet{martig16} calibration underestimates [C/N]-based ages in the older regime by up to a factor of two when compared with asteroseismic ages.
In a parallel paper, \citet{ness16} used \emph{The Cannon} \citep{ness15} to extract mass and age information for $\sim$50,000 stars from the APOGEE DR12 spectra from spectral regions with CN absorption lines.
Subsequent works extended the [C/N]-age relations to open clusters observed in \emph{Gaia}-ESO iDR5 and APOGEE DR14 \citep{casali19} and in APOGEE DR17 \citep{spoo22}. \citet{jofre21} used this chemical clock as a tracer of the two different disc sequences in the age-metallicity relation.
More recently, \citet{roberts25} employed the APOKASC-3 sample \citep{pinsonneault25} to investigate the [C/N]-age relationship, while \citet{grazina25} analysed 44 open clusters from the \emph{Gaia}-ESO survey \citep{randich22}, and \citet{spoo25} studied four globular clusters from APOGEE DR17 \citep{apogeedr17}. Finally, \citet{pakstiene26} and \citet{curjuric26} used, respectively, 1,250 TESS giant stars and 28 open clusters from Stellar Population Astrophysics (SPA) observing programme \citep{origlia19} to investigate [C/N] as an age indicator.

In this work, we aim to calibrate the chemical clock [C/N] using a sample of \kepler giant stars \citep[present in][]{willett25} with asteroseismic ages estimated using the global seismic parameters, \numax and $\Delta\nu$, and abundances from Apache Point Observatory Galactic Evolution Experiment (APOGEE) DR17 \citep{apogeedr17}. The empirical relationship deduced from this calibration sample will be applied to other seismic catalogues with APOGEE abundances, e.g. K2 and TESS from \cite{willett25}, and a sample of open clusters present in the APOGEE DR17 and \emph{Gaia}-ESO surveys to validate and constrain the relationship. We also investigate the limits of this relationship and we compare it with other chemical clocks relationships based on Galactic evolution.

%\SB{This is a personal preference, but I always find it useful when figuring out if a paper is interesting enough to continue reading (aka do the authors actually have one or more hypotheses/objectives: I always put the specific aim / questions of a paper at the very end of the introduction. So for this paper, I would switch "In this work ..." and "Previous studies have ..." and also highlight your aims by listing questions / objectives iteratively (that way people who speed-read will very quickly know what's in your paper.)}

The paper is composed as follows. In Sec.~\ref{sec:datasample}, we describe the data samples used in this work. In Sec.~\ref{sec:relationship}, we obtain the age-[C/N]-[Fe/H] relationship for the calibration sample and we apply it to the validation samples. In Sec.~\ref{sec:validityandlimits}, we discuss the validity and limitations of this relationship. In Sec.~\ref{sec:comparisoncc}, the comparison among [C/N] and other chemical clocks is shown. In Sec.~\ref{sec:application}, the relationship is applied to red giants in the APOGEE survey. In Sec.~\ref{sec:conclusions}, we summarise and conclude.

\section{Data samples}
\label{sec:datasample}
Our dataset consists of a calibration sample -- \kepler giant stars -- for which we will derive the [C/N]-age-[Fe/H] relationships. We will then apply these relationships to other samples -- used as validation samples -- including open clusters, TESS, and K2 targets, in order to assess the reliability of our calibration. These samples are described below, with quality cuts further discussed in Sec.~\ref{sec:qualitycuts}.

\subsection{Calibration sample: \kepler data set}

Our calibration sample consists of 10,170 %9,730 
giant stars observed by the \kepler mission \citep{Borucki2010, gilliland10} and presented in \cite{willett25}. For these stars, astrometric constraints are taken from \gaia DR3 \citep{gaiadr3}, and chemical abundances are drawn from the seventeenth data release of the APOGEE survey \citep[DR17;][]{apogeedr17}, which provides high-resolution near-infrared spectra and precise elemental measurements.
The atmospheric parameters and chemical abundances used in this work are derived using the standard APOGEE data analysis pipeline -- the APOGEE Stellar Parameters and Chemical Abundances Pipeline \citep[ASPCAP;][]{garcia2016}. %A full description of the pipeline is provided by Holtzman et al. (in preparation) \mkn{I think this paper now exists?}. 
Carbon and nitrogen abundances are determined through a combined analysis of their contributions to numerous molecular bands, including CN and CO, as described in \citet{smith21}.
Stellar ages are determined through asteroseismology. The nearly continuous 4.5-year \kepler photometric time series provides exceptionally high signal-to-noise oscillation data, enabling the precise detection and characterisation of solar-like oscillation modes in giant stars. This allows for robust derivations of stellar masses and radii, which in turn yield age estimates with typical uncertainties of less than 20\%. Stellar parameters (mass, radius, age, and distance) are inferred using the PARAM code \citep{dasilva2006, rodrigues2017}, which performs a Bayesian comparison between observational data and stellar models. The resulting catalogue is described in \citet{willett25}.
The combination of reliable stellar ages and high-precision chemical abundances makes this dataset particularly well-suited for calibrating empirical relationships between [C/N] and stellar age.
%Thanks to asteroseismology, we have also information on the evolutionary stage of these stars. They are composed of XXX lower red giant branch (RGB) stars ($2.5 <= \log~g <= 3.5$), XXX upper RGB stars ($\log~g > 2.5$) and XXX red clump stars (RC).

\subsection{Validation samples}
Since the fits for the [C/N]-age-[Fe/H] relationships are performed on a high-quality calibration sample, composed of stars located in a single field of the Galactic disc, we test the fits on different validation samples to verify that they can be applied to other Galactic fields and to data based on different abundance sets and/or stellar ages.
The validation samples consist of open clusters presented in \citet{spoo22} and \citet{grazina25}, as well as two asteroseismic samples: K2 and TESS targets from \citet{willett25}.

\subsubsection{Open clusters}
This sample consists of open clusters from the Open Cluster Chemical Abundances and Mapping survey \citep[OCCAM,][]{occam}, based on APOGEE Data Release 17 \citep[DR17,][]{apogeedr17}, and from the \emph{Gaia}-ESO survey \citep{randich22}. The mean [C/N] ratios for the OCCAM clusters are provided in \citet{spoo22}. In that work, cluster member stars are selected based on reliable C and N measurements (using ASPCAP bitwise flags) and a membership probability greater than 70\% in radial velocity, proper motion, and metallicity. Stars with $\log~g > 3.3$ and red clump stars are identified using the flags from the catalogue by \citet{bovy14} and Bovy et al. (in prep). From this sample, we retain only clusters with at least three member giants, resulting in a final sample of 42 clusters.
For the \emph{Gaia}-ESO clusters, membership probabilities are taken from \citet{huntreffert24} and \citet{jackson22}, while the mean [C/N] for giant members are shown in \citet{grazina25}. As with the OCCAM sample, we select only clusters with at least three member giants, resulting in a final sample of 37 clusters.
Since these open clusters come from two different spectroscopic surveys -- using different instruments, models, tools, and wavelength ranges -- it is necessary to bring all [C/N] measurements onto a consistent scale. We compute the offset using stars in common between the two surveys, finding a systematic offset of 0.05 dex. This offset is then applied to place the \emph{Gaia}-ESO [C/N] abundances onto the APOGEE scale. We also verify that the C and N abundance differences between the two surveys show no significant trends with C and N abundances or atmospheric parameters.
Finally, the ages of the star clusters used in this work are taken from \citet{cantat20}, as also used by \citet{spoo22} and \citet{grazina25}.

\subsubsection{Asteroseismic samples}
The catalogues based on asteroseismic observations from the K2 \citep{howell14} and TESS \citep{ricker14} missions are combined with astrometric parameters from \emph{Gaia} DR3 \citep{gaiadr3} and spectroscopic measurements from APOGEE DR17. Stellar masses, radii, and ages are computed using the PARAM code \citep{dasilva2006, rodrigues2017}. These catalogues are described in detail in \citet{willett25}.
The TESS sample consists of 1,921 targets observed during the mission's first year of operations in its southern continuous viewing zone \citep[SCVZ; see][for more details]{mackereth21}.
The K2 dataset includes 10,016 targets,
%6,155 targets with reliable ages, 
observed across 20 campaigns along the ecliptic plane. Unlike the original \kepler and TESS missions, K2 observed stars spanning a wide range of Galactocentric radii.

\subsection{Selection criteria}
\label{sec:qualitycuts}
For each sample in this study that includes data from \emph{Gaia} and APOGEE, we apply a series of quality cuts to ensure reliable measurements. We exclude stars with \texttt{RUWE} > 1.4 and those flagged with any of the following in the \texttt{ASPCAPFLAG}: \texttt{STAR\_BAD, STAR\_WARN, C\_M\_BAD}, or \texttt{N\_M\_BAD}. We also remove stars flagged with \texttt{BAD\_PIXEL, VERY\_BRIGHT\_NEIGHBOR, LOW\_SNR}, or \texttt{PERSIST\_HIGH} in the \texttt{STARFLAG}, and we discard stars with relevant \texttt{ELEMFLAG} values set for either carbon or nitrogen.
Regarding the asteroseismic parameters, we apply the flags shown in \citet{willett25} for bad values: \texttt{warn\_low\_numax, warn\_high\_Dnu, warn\_nveAv}, and  \texttt{warn\_massDiff} for K2 only due to unreliable \dnu determinations for core helium burning stars.
In addition, we restrict the sample to stars with [Fe/H] > –1, [C/N] uncertainties < 0.2 dex, and [C/N] > –0.8, since stars below this threshold are likely results of non-standard evolution. Finally, outliers were identified using the DBSCAN clustering algorithm in the [C/N]-log(Age/yr) plane and removed from the sample to ensure a robust fit. %Finally, outliers in [C/N]-log(Age) relation are identified using a $3-\sigma$ clipping and excluded from the analysis to ensure a robust fit.

\begin{figure*}
\centering
\includegraphics[scale=0.45]{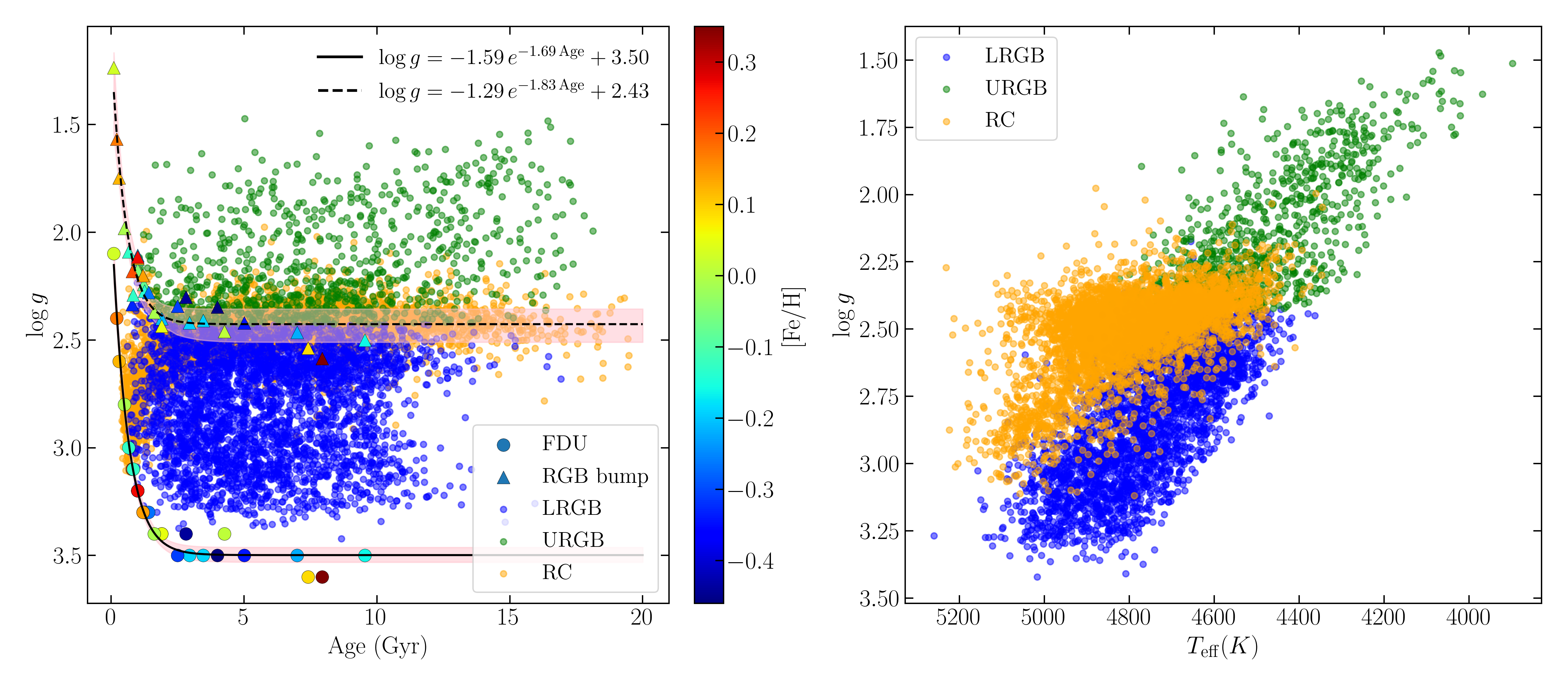}
\caption{{\it Left panel:} The corresponding theoretical \logg at the FDU (circles) and RGB bump (triangles) as a function of age and metallicity for a set of PISA isochrones presented in \citet{casali19}. The symbols are colour-coded by their metallicities. The lines and the shaded areas represent the fits and the 3$\sigma$ intervals, respectively. \kepler stars are overplotted with different colours relating their evolutionary stage. {\it Right panel:} The Kiel diagram for the \kepler sample. The three different colours represent lower-RGB (LRGB, blue), upper-RGB (URGB, green), red clump (RC, orange) stars, respectively. %\SB{It looks like there should be enough space to spell out the abbreviations in the legend (so other people can easily take a screenshot and add it to a presentation without needing annotations).} 
\label{fig:kepler_kieldiagram}}
\end{figure*}

All selected samples include stars in different evolutionary stages. We classify them into three categories: lower red giant branch (LRGB, defined as RGB stars before the RGB luminosity bump), upper red giant branch (URGB, RGB stars after the RGB luminosity bump), and red clump (RC) stars.
For \kepler stars, the evolutionary classification is available from the period spacing of dipole mixed modes via the "EvoState" column in the catalogue, where RGB stars are flagged as 1 and RC stars as 2.
However, a distinction between lower and upper RGB is not provided. 
We distinguish between lower and upper RGB stars by comparing the surface gravity ($\log g$) of each star with theoretical predictions for the first dredge-up (FDU) and the RGB luminosity bump (hereafter RGB-bump). These reference values are derived from a set of PISA stellar isochrones \citep{dellomodarme12,tognelli18}, computed over a range of ages and metallicities (see Fig.~\ref{fig:kepler_kieldiagram}, left panel). The theoretical $\log g$ values at the FDU and RGB-bump are taken from \citet{casali19}, where they are provided as a function of stellar age and metallicity (circles and triangles in Fig.~\ref{fig:kepler_kieldiagram}, respectively) and subsequently fitted with polynomial relations.
We then use these polynomial fits to define evolutionary regions in surface gravity: stars located between the FDU and RGB-bump loci are classified as lower RGB (LRGB), while stars with $\log g$ values lower than the RGB-bump prediction are classified as upper RGB (URGB). 
This separation converges to nearly constant values of surface gravity, approximately $\log g \simeq 3.5$ dex for the FDU and $\log g \simeq 2.4$ dex for the RGB-bump, for ages older than $\sim 2.5$ Gyr. At younger ages, however, the location of both the FDU and the RGB-bump becomes strongly dependent on age, with significant variations in their $\log g$ values across the explored parameter space.
Considering the LRGB and URGB separation described above, together with the RC classification based on period spacing, the corresponding distribution in the Kiel diagram for the \kepler sample is shown in Fig.~\ref{fig:kepler_kieldiagram} (right panel). 

For K2 and TESS, no measurements of the period spacing of dipole mixed modes are available in \citet{willett25}, which precludes an asteroseismic distinction between RGB and RC stars. We therefore identify RC stars using the APOGEE DR17 red clump catalogue \citep[][Bovy et al. in prep.]{bovy14}, and classify all remaining giants as RGB stars, which are then subdivided into LRGB and URGB using the empirical fits described above. In those samples, we cannot exclude the presence of early-AGB type stars among the URGB stars, since their identification is extremely challenging without the asteroseismic information about their evolutionary stage. We might expect deviations in [C/N] for those stars relative to the bulk of the URGB population. However, stars with large deviation in [C/N] are already identified as outliers by the DBSCAN algorithm and removed during the cleaning procedure, making the initial inclusion of such objects non-critical for the final sample. %ANDREA'S COMMENTS: isn't URGB a mix of RGB and eAGB? if this is the case then you may expect a significant scatter in [C/N] vs mass / age, assuming we do see post bump extra mixing in this sample

Regarding the sample of open clusters, the division of cluster members in the three categories is already provided by \citet{spoo22} and \citet{grazina25}.\\
Taking into account all quality cuts, the \kepler sample comprises 9,122 stars: 3,424 LRGB, 910 URGB, and 4,788 RC stars, while the TESS and K2 samples include 526, 522, and 528 stars, and 3,586, 1,089, and 1,381 stars, respectively.

\begin{figure*}
\centering
\includegraphics[scale=0.42]{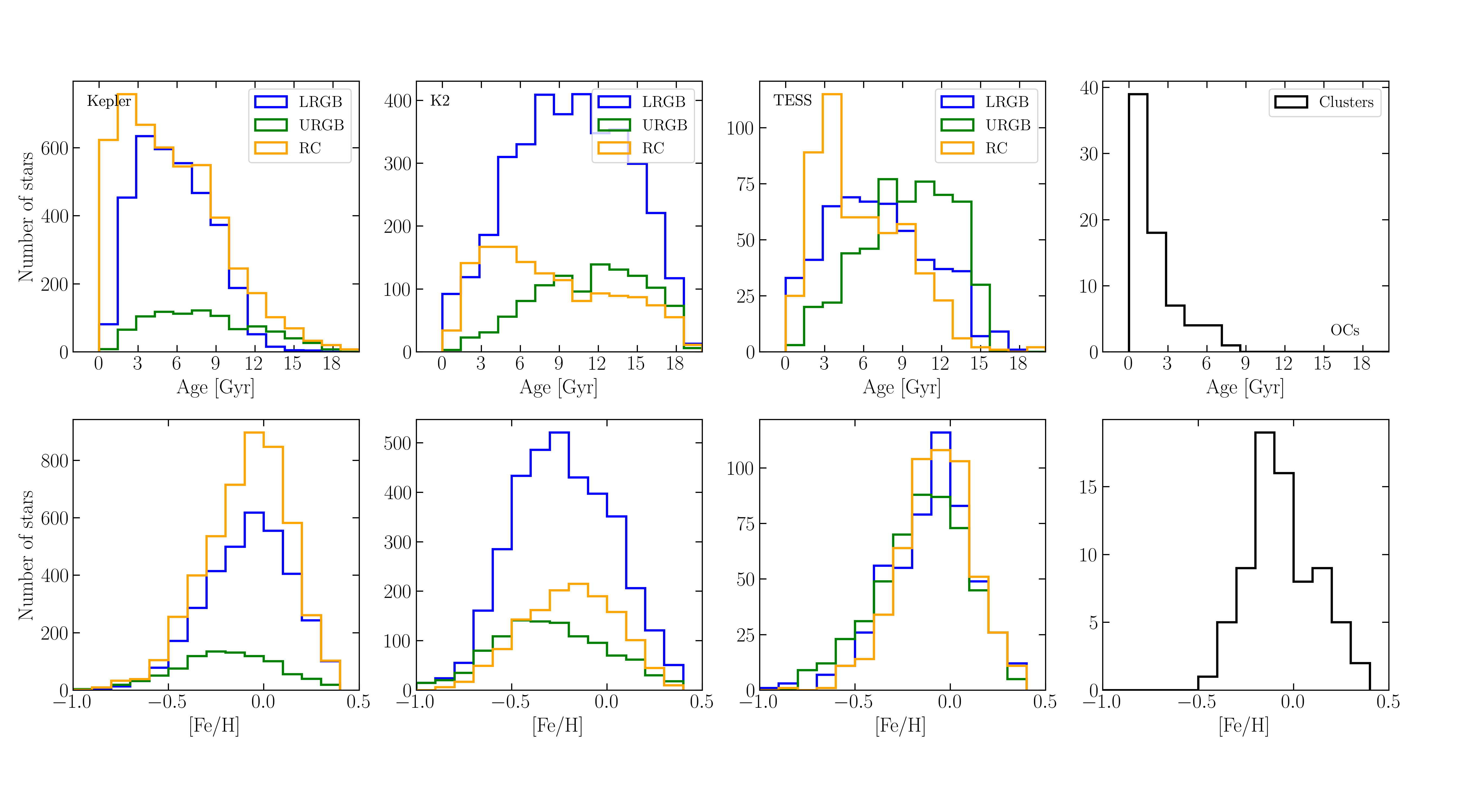}
\caption{Age (top panels) and metallicity (bottom panels) distributions for the four data samples (\emph{Kepler}, K2, TESS, open clusters), shown separately for the three subsets: lower-RGB (LRGB, blue), upper-RGB (URGB, green), and red clump (RC, orange) stars. %\SB{Spell out the abbreviations at least on the first and last panel? And maybe increase the fontsize of the legend to be at least the fontsize of the caption?} 
\label{fig:hist}}
\end{figure*}

Figure~\ref{fig:hist} shows the age and metallicity distributions for each sample. In all asteroseismic samples, RC stars exhibit distributions skewed toward younger ages compared to both LRGB and URGB stars. The LRGB distribution peaks at approximately 5-6 Gyr for the \kepler and TESS samples, and at slightly older ages for the K2 sample.
URGB stars show age distributions peaking at older ages in the K2 and TESS samples, whereas in the \kepler sample the distribution is more weighted toward younger ages. We do not exclude the possibility that these differences arise from selection effects affecting the age distributions.
Finally, open clusters exhibit systematically younger ages than field stars, with typical cluster ages below 7 Gyr. This is expected due to the survival time of open clusters \citep[e.g.,][]{viscasillas23}. Regarding metallicity, all datasets (particularly for LRGB and RC) show distributions peaked around solar metallicity, with the exception of K2, which is slightly shifted toward sub-solar values. For the URGB stars in both \kepler and K2, the distributions peak at sub-solar metallicity, while URGB stars in TESS show a similar distribution for LRGB and RC stars. Finally, open clusters show a metallicity distribution peaked at solar values, with a narrower range of $[-0.5, 0.4]$~dex.

\section{Age-[C/N]-[Fe/H] relationship}
\label{sec:relationship}
We calibrate a relationship between the asteroseismic stellar age, [C/N] and [Fe/H] using the \kepler sample for the three subsets: LRGB, URGB and RC. Giants could be affected by extra mixing, which becomes a non-negligible effect for stars that have passed the RGB bump and have metallicities lower than $\rm [Fe/H] = -0.4$ dex \citep{shetrone19}. %For this reason, we adopt a more conservative approach and derive the relationship exclusively for LRGB stars.
As a consequence, we place greater confidence in LRGB stars and in the relationships derived from them. In contrast, URGB and RC stars are considered less reliable in this context because of extra mixing effects. Nevertheless, we also derive relationships for these evolutionary stages, restricting the sample to stars with $\rm [Fe/H] > -0.4$ for URGB and RC stars, following \cite{shetrone19}.
Since the \kepler sample contains few stars in the youngest regime, we include open clusters from the APOGEE and \emph{Gaia}-ESO surveys in the fit to better constrain this regime.
We use the same functional form as proposed by \citet{roberts24} for stellar mass and \citet{roberts25} for stellar age. This relationship is as follows:

\begin{equation}
\begin{split}
\log(\mathrm{Age/yr}) =\ & c_{5}\mathrm{[C/N]}^2 + c_{4}\mathrm{[C/N]} + c_{3}\mathrm{[Fe/H]}^2 \\
&+ c_{2}\mathrm{[Fe/H]} + c_{1}\mathrm{[C/N]}\mathrm{[Fe/H]} + c_{0}
\end{split}
\label{eq:1}
\end{equation}

%Unlike \citet{roberts25}, we exclude stars with [C/N] > 0, as the [C/N]–age trend is completely flat in this regime. For such stars, it is not possible to estimate a precise age, but we can just say that the star is likely old. The selected validation sample is shown in Fig.~\ref{fig:validationsample}. 
We fit this quadratic relationship to the full \kepler sample ($\rm [Fe/H] > -1$), the LRGB sample ($\rm [Fe/H] > -1$), the URGB sample ($\rm [Fe/H] > -0.4$), and the RC sample ($\rm [Fe/H] > -0.4$). 

%\SB{I have not found a specific part where to ask these questions: 1) is there a reason why you only use ages and not also masses (I may be wrongly assuming that this should be another part of the asteroseismic results if you fit age anyways. While the mass-fits are of course not directly applicable to Galactic archaeology, I think there is a lot of follow-up research possible (and also I would be interested in seeing some effects of [C/N] with mass, especially for primary and secondary red clump). That said: I have read nothing about the primary and secondary red clump in your paper. Have you by any chance looked into this and if you can for example achieve a better fit when separating them (as the different Teff/luminosity regimes they inhabit may also cause some differences in {[C/N]} behaviour)?}

\begin{figure*}
\centering
\includegraphics[scale=0.5]{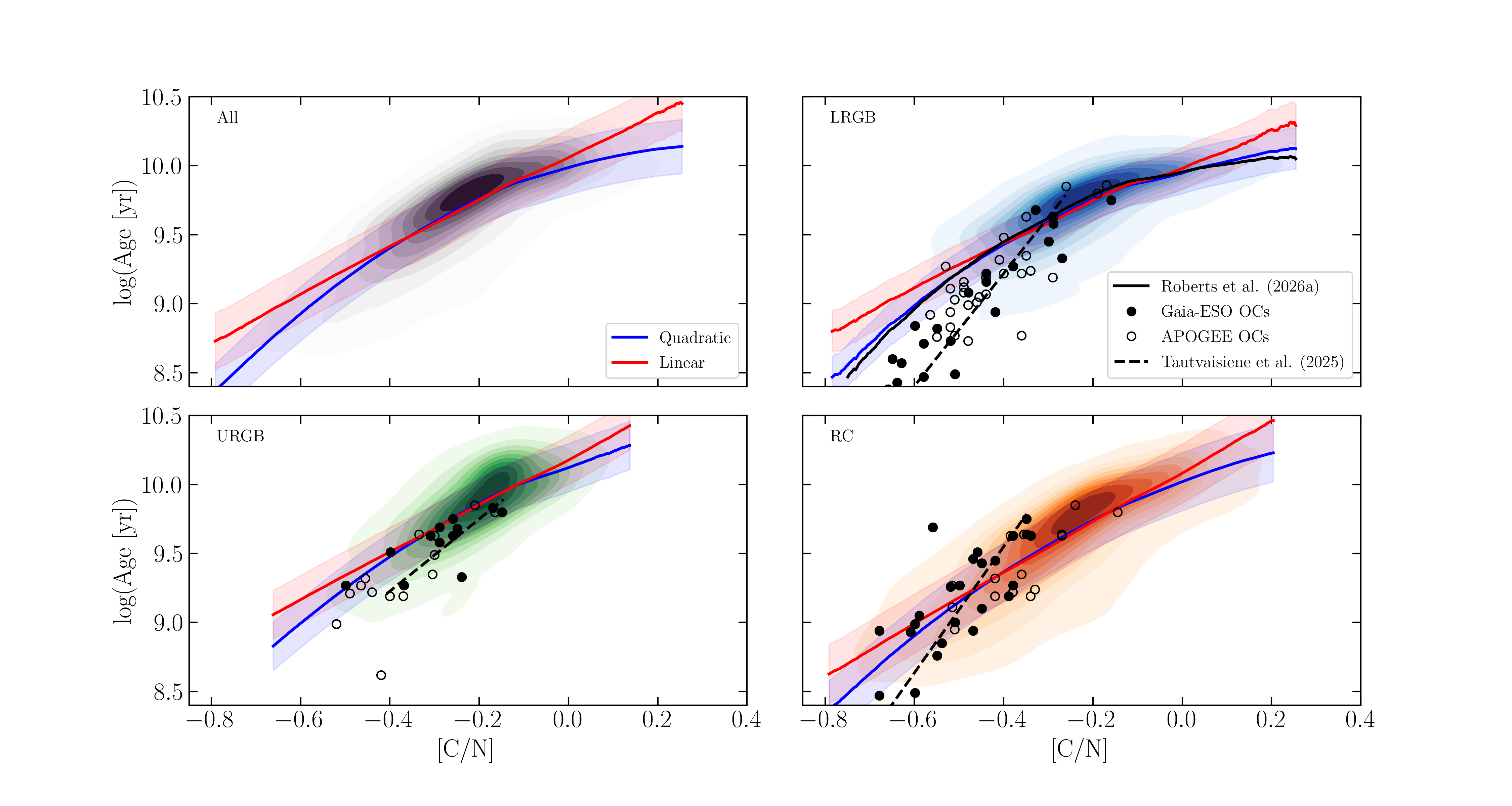}
\caption{log(Age[yr]) vs [C/N] panels for all \kepler sample and its subsets: LRGB, URGB, RC. To illustrate the multivariate fit, including its dependence on metallicity, we binned the data in [C/N] and computed the mean [Fe/H] within a sliding window around each bin. This yields a smoothed representation of the metallicity as a function of [C/N]. The fitted relationships were then evaluated using these mean [Fe/H] values, effectively projecting the multivariate model onto the [C/N] axis. In addition, open clusters from APOGEE and \emph{Gaia}-ESO survey, with the offset with respect to the APOGEE abundances applied to the latter, and relations from \citet{roberts25} and \citet{grazina25} are shown. %\SB{Again, I think you have enough space to spell out R25, GES, and T25 as well as "Quad". Also: I do not really see a clear trend of {[Fe/H]} in the scatter points. Would that arise when using medians in bins? Otherwise I am not sure if the color-dimension is helpful here (instead you could maybe switch again to your previous yellow/blue/black one. With that many points, I am also wondering if it would be better to move to a hist2d map or even contour lines (scatter points never allow to see the density distribution, once points are overlapping.)} 
\label{fig:validationsample}}
\end{figure*}

We also fit a linear relationship to each subsample for comparison with the quadratic relationship and with previous results in the literature:

\begin{equation}
\log(\mathrm{Age/yr}) = c_{4}\mathrm{[C/N]} + c_{2}\mathrm{[Fe/H]} +  c_{0}
\label{eq:2}
\end{equation}

%The $R^2_{adj}$ is slightly better if we use the quadratic relationship. For this reason, we will use this one for this work.

The fit parameters are shown in Table~\ref{tab:fit} for the \kepler sample, both including and excluding the open clusters from the regression. The adjusted coefficient of determination, $R^2_{\mathrm{adj}}$, is always larger than 0.6, except for URGB stars. This may be due to the smaller number of URGB stars compared to LRGB and RC, as well as their limited coverage in log(Age/yr) and [C/N] (i.e. they do not span values below $-0.5$ or above $0$ dex).
The quadratic relationships have lower Bayesian and Akaike information criteria (BIC/AIC) than the linear ones, indicating a better balance between goodness of fit and model complexity. The differences, $\Delta$BIC/AIC, are larger than 10 for all relationships except for the URGB sample, for which $\Delta \rm BIC/AIC \sim 2$.

In Fig.~\ref{fig:validationsample}, we compare the linear and quadratic fits for the full sample and the corresponding subsamples. To effectively visualise the multivariate models as a two-dimensional projection of $\log(\mathrm{Age/yr})$ vs [C/N], we applied a rolling mean to determine the expected [Fe/H] as a function of [C/N].
In the first panel, we show the full \kepler sample with the corresponding linear and quadratic relationships. The other panels display the two different relationships, along with a comparison to the relationships from \citet{roberts25} (restricted to LRGB stars) and \citet{grazina25}. As shown in the figure, the quadratic relationships derived in this work and by \citet{roberts25} are broadly consistent, with slight discrepancies at young and old ages. This is partly due to the lack of stars in these regimes, which also reduces the precision of the calibration, and to an offset between the two age estimates in the older regime, where APOKASC-3 shows younger ages than \citet{willett25}, as already pointed out by \citet{casali25}.
However, the difference with \citet{grazina25} is substantial because their calibration is based on open clusters, which exhibit a different slope than field stars. This is due to their sparsity in the old age regime ($\log(\mathrm{Age/yr}) > 9.8$) covered by the \kepler\ sample, as well as their broader extension in the young regime. 

\begin{table*}
 \caption{Parameters of the fitting using Eq.~\ref{eq:1} (Quad) and Eq.~\ref{eq:2} (Lin), respectively. }
\begin{center}
\begin{tabular}{ccccccccc}
\hline
Sample & Eq. & $c_{0}$ & $c_{1}$ & $c_{2}$ & $c_{3}$ & $c_{4}$ & $c_{5}$ & $R^{2}_{adj}$ \\
\hline
%LRGB & Quadratic & 10.125  &  0.136 & 0.466 &  0.228 & 1.852 &  0.435 & 0.59 \\
%LRGB & Linear & 10.108 &  --      & 0.397 &  --      & 1.626 &  --    & 0.58 \\
 All  (\emph{Kepler}) & Quad       & 10.057 &  0.381    & 0.366 & 0.446 & 1.118 & $-$1.284 &  0.64  \\
 LRGB (\emph{Kepler}) & Quad       & 10.064 & $-$0.027  & 0.415 & 0.172 & 1.295 & $-$0.601 &  0.61  \\
 URGB (\emph{Kepler}) & Quad       & 10.184 &  0.610    & 0.315 & 0.112 & 1.399 & $-$0.771 &  0.51  \\
 RC   (\emph{Kepler}) & Quad       & 10.042 & $-$0.047  & 0.187 & 0.403 & 1.252 & $-$1.096 &  0.68  \\
 All  (\emph{Kepler}) & Lin        & 10.124 &  --     & 0.211 & --      & 1.768 & --      &  0.64  \\
 LRGB (\emph{Kepler}) & Lin        & 10.100 &  --     & 0.412 & --      & 1.598 & --      &  0.61  \\
 URGB (\emph{Kepler}) & Lin        & 10.204 &  --     & 0.197 & --      & 1.679 & --      &  0.51  \\
 RC   (\emph{Kepler}) & Lin        & 10.116 &  --     & 0.180 & --      & 1.882 & --      &  0.66  \\
 LRGB (\kepler + OCs) & Quad  & 10.040  & $-$0.393 & 0.338 & 0.142  & 1.010 & $-$1.285 &   0.61  \\
 URGB (\kepler + OCs) & Quad  & 10.181  &    0.648 & 0.320 & 0.130  & 1.341 & $-$1.056 &   0.51  \\
 RC   (\kepler + OCs) & Quad  & 10.042  & $-$0.016 & 0.193 & 0.414  & 1.249 & $-$1.104 &   0.68  \\
 LRGB (\kepler + OCs) & Lin   & 10.118 &  --     & 0.440 & --     & 1.690 &  --     &   0.61  \\
 URGB (\kepler + OCs) & Lin   & 10.215 &  --     & 0.195 & --     & 1.762 &  --     &   0.51  \\
 RC   (\kepler + OCs) & Lin   & 10.117 &  --     & 0.178 & --     & 1.887 &  --     &   0.66  \\
\hline
\end{tabular}
\end{center}
\label{tab:fit}
\end{table*}

\begin{figure*}
\centering
\includegraphics[scale=0.5]{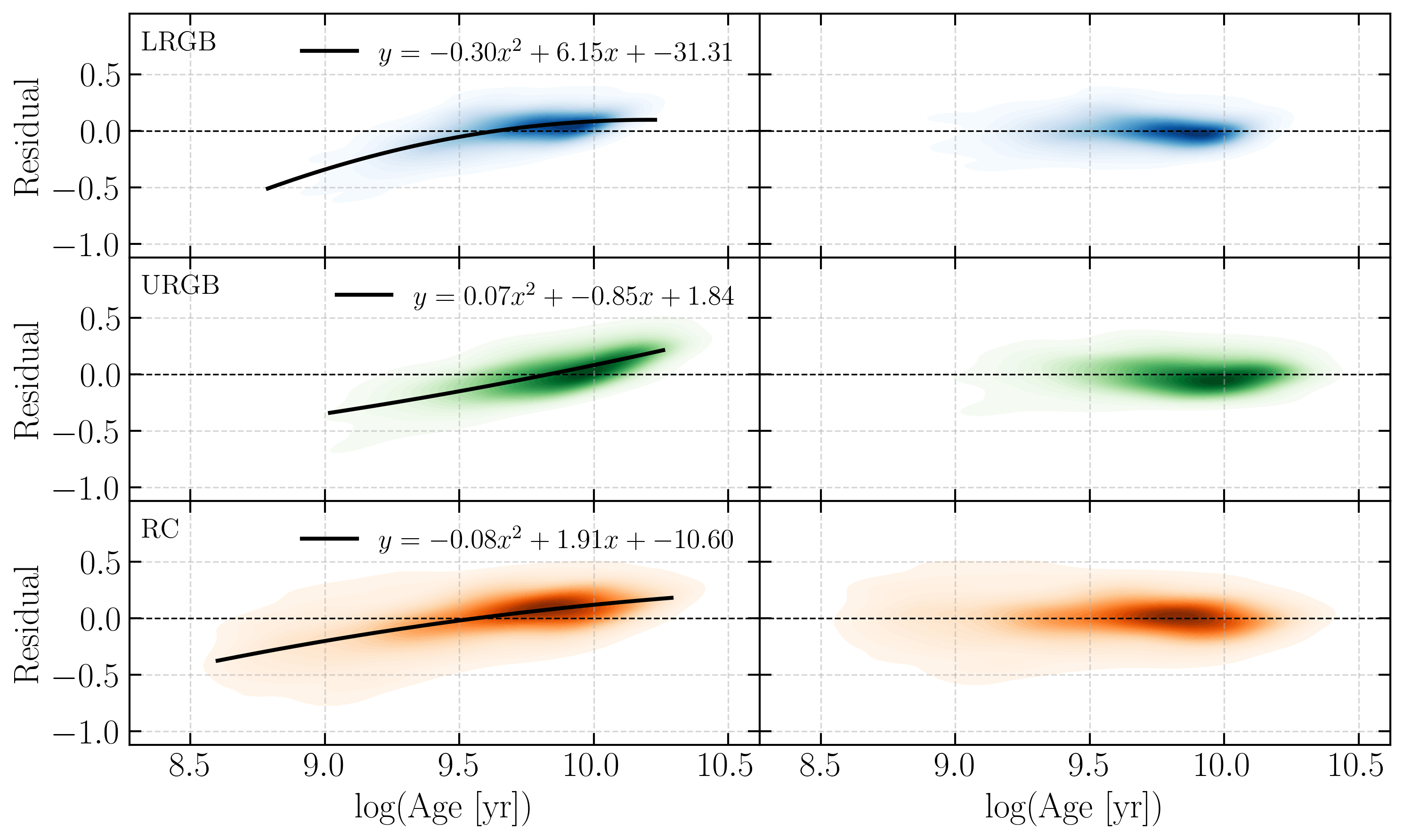} %{Kepler_residuals_corr_vs_nocorr2.png}
\caption{The residuals $\log(\mathrm{Age_{seismic}}) - \log(\mathrm{Age_{[C/N]}})$ are shown for each \kepler subsample (LRGB, URGB, RC). The left panels show the uncorrected residuals together with the corresponding fits (the fit parameters are reported in the figure). The right panels show the corrected residuals, obtained by applying the relations illustrated in the left panels. %\mkn{it would be good to understand what the expectation of this residual scatter is from uncertainty alone (quadrature sum of median uncertainty of both). If the scatter around the line is not larger than that these paraemters fully explain the data to within uncertainties. If the scatter exceeds that value, there must either be an unaccounted for parameter or stochasticity (some random effect on top of the systematic) that does not enable a perfect reconstruction. That is useful to know if there are any additional correlations with the residuals if they exceed uncertainty expectation. If the scatter exceeds sqrt(error age seismic squared plus error age c/n squared) you might plot residual versus the parameters of your model (these should be flat) but that is a good verification step, as well as versus other available variables to check for any correlation e.g. colour, [X/H] for all other elements X, microturbulence, velocity scatter in APOGEE (e.g. are things with larger velocity scatter which may be binaries the things with highest residuals?)}. \SB{I think it would be good to move to hist2d or contourplots (or the corner-like combo of contour+scatter) to give an idea of the density distribution of points).} 
\label{fig:residuals}}
\end{figure*}

We apply the relationships defined in Eq.~\ref{eq:1} -- with OCs included in the regression -- to the \kepler stars and compare the input seismic ages with the resulting chemical ages, derived from the [C/N] and [Fe/H] values, in each subsample (see Fig.~\ref{fig:residuals}). The residuals between these two age estimates indicate that the relationships are less reliable for URGB and RC stars than for LRGB stars. In more detail, the relationship for LRGB stars becomes less accurate for $\log(\rm Age/yr) \lesssim 9.3$, corresponding to linear ages younger than $\sim 2$ Gyr, as expected for the reasons discussed in Sec.~\ref{sec:limits}. The age in this regime tends to be overestimated by a factor of up to $1.03$ ($1.9$ in linear age) with respect to the seismic ones, with a $\sigma \sim 0.2$ in the residual. There is also a discrepancy for ages older than $\log(\rm Age/yr) \gtrsim 10$, corresponding to linear ages older than $\sim 10$ Gyr. The age in this regime tends to be underestimated by a factor of $1.01$ ($1.3$ in linear age), with a $\sigma \sim 0.1$ in the residual.
Regarding the URGB sample, the residuals show a steep increasing trend with age. In this case, the smaller sample size and the narrower distribution in both age and [C/N] with respect to LRGB and RC subsets must be taken into account, as they may bias the resulting residuals and do not allow robust conclusions.
Regarding the RC sample, the number of stars is larger, and the sample extends to younger ages compared to the LRGB and URGB subsets. RC stars tend to overestimate young ages by a slightly smaller factor than LRGB ($\sim 1.02$, 1.8 in linear age), but with a larger scatter ($\sigma \sim 0.25$). In the old-age regime, they underestimate seismic ages by a factor similar to LRGB, but again with a larger scatter.
Overall, LRGB stars provide more reliable constraints at older ages, whereas RC stars are better suited to probing younger ages than LRGB stars.

Moreover, the residual scatter is larger than that expected from the uncertainties alone. To investigate its origin, we also examine possible correlations between the residuals and additional stellar parameters, including $T_{\mathrm{eff}}$, $\log g$, [$\alpha$/Fe] and $v \sin i$, but no significant trends are found. This suggests that either additional unaccounted-for physical parameters or intrinsic stochasticity contribute to the observed dispersion in the [C/N]-age-[Fe/H] relationship.

To account for the deviations from the input ages in each subsets, we model the residuals with a polynomial function and derive a correction that improves the chemical age estimates in the regime where the relationships are less reliable. The polynomial functions are reported in Fig.~\ref{fig:residuals} and only depend on the stellar age. The right panels of Fig.~\ref{fig:residuals} show the residuals if corrections are applied.
These corrections can only be applied when an independent age estimate is available in addition to that derived from [C/N] (e.g., an age estimate from asteroseismology). In any case, when using Eq.~\ref{eq:1} to estimate the ages of stars for which no independent age constraints are available, it should be kept in mind that ages below 
$\log(\rm Age/yr) \sim 9.3$ (and above $\log(\rm Age/yr) \sim 10$) tend to be overestimated (and underestimated) and should therefore be used with caution. However, the [C/N]-based ages still allow a robust relative classification, in the sense that stars with ages below $\sim 2$ Gyr can be considered young, while those above $\sim 10$ Gyr can be safely regarded as old.

%Figure~\ref{fig:residuals_corr} shows the residuals after applying the correction derived from the polynomial function.

%\begin{equation}
%\begin{split}
%\log(\mathrm{Age_{seismic}}) - \log(\mathrm{Age_{[C/N]}}) = \\  
%a \cdot \log(\mathrm{Age_{seismic}})^2 + b \cdot \log(\mathrm{Age_{seismic}}) + c
%\end{split}
%\end{equation}

\subsection{Application to the validation samples}
\begin{figure*}
\centering
\includegraphics[scale=0.5]{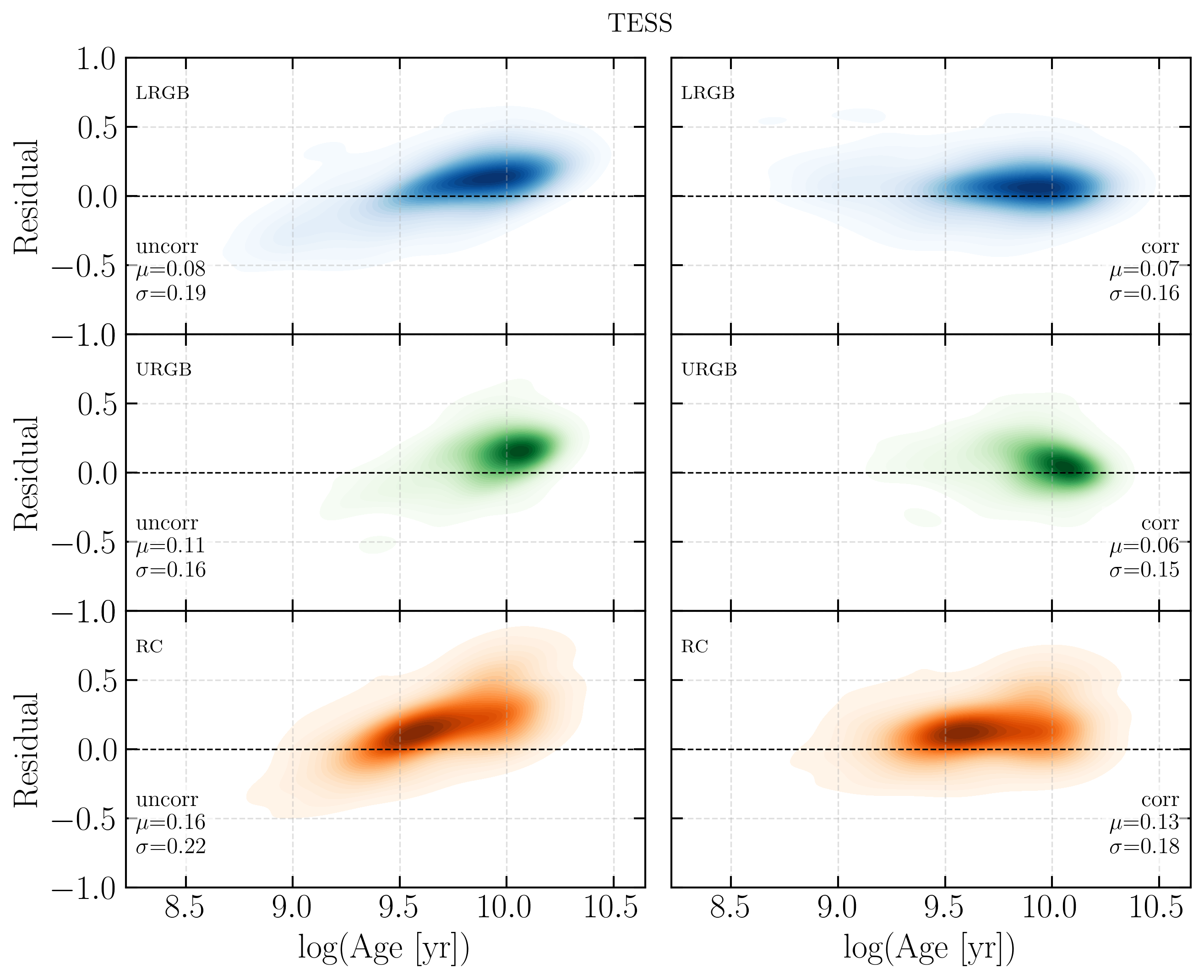}%{tess_residuals_corr_vs_nocorr.png}
\caption{The residuals $\log(\mathrm{Age_{seismic}}) - \log(\mathrm{Age_{[C/N]}})$ vs $\log(\mathrm{Age_{seismic}})$ are shown for each TESS subsample (LRGB, URGB, RC). The left panels are the uncorrected residuals, the right panels are the corrected ones. \label{fig:residuals_tess}}
\end{figure*}

\begin{figure*}
\centering
\includegraphics[scale=0.5]{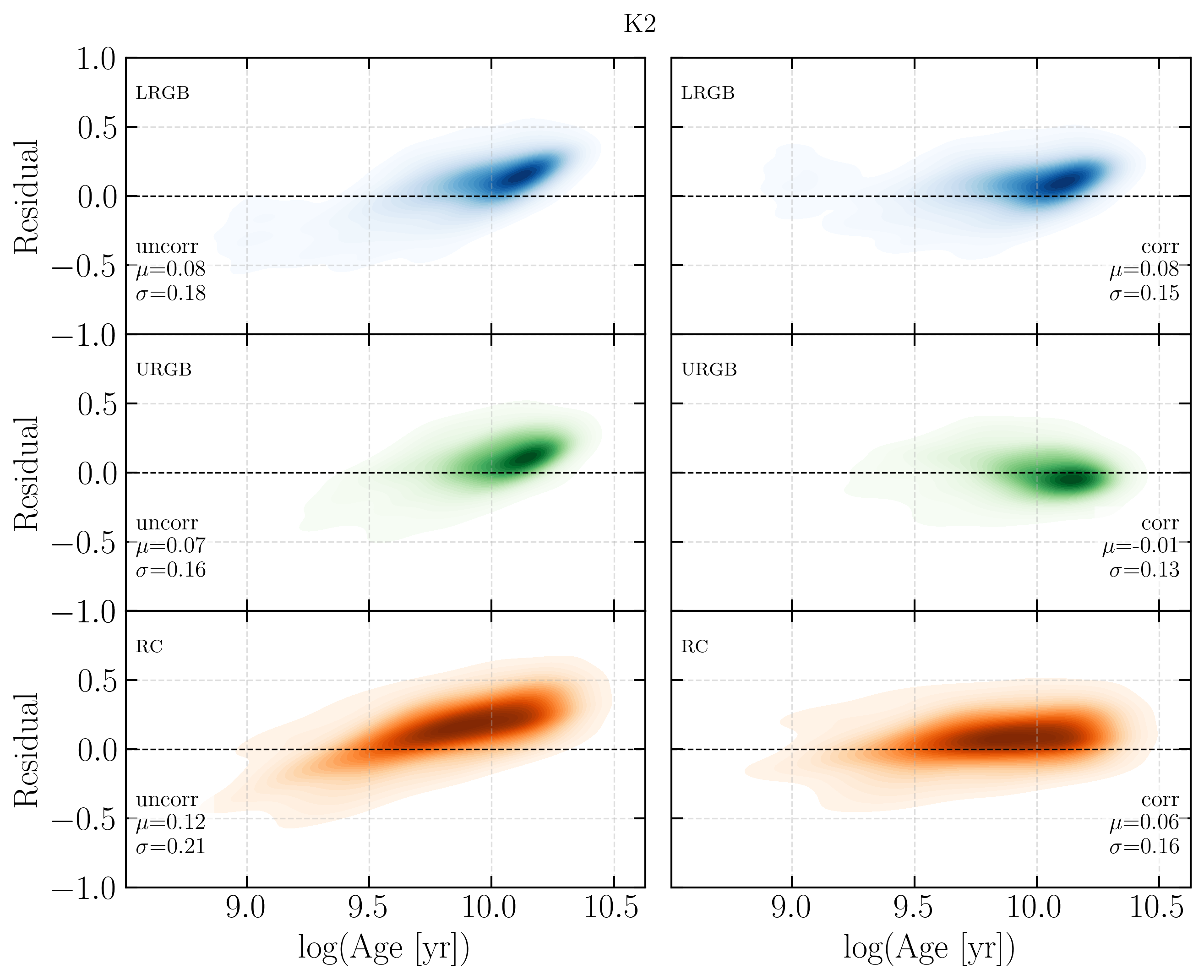}%{k2_residuals_corr_vs_nocorr.png}
\caption{The residuals $\log(\mathrm{Age_{seismic}}) - \log(\mathrm{Age_{[C/N]}})$ vs $\log(\mathrm{Age_{seismic}})$ are shown for each K2 subsample (LRGB, URGB, RC). The left panels are the uncorrected residuals, the right panels are the corrected ones. \label{fig:residuals_k2}}
\end{figure*}

\begin{figure*}
\centering
\includegraphics[scale=0.5]{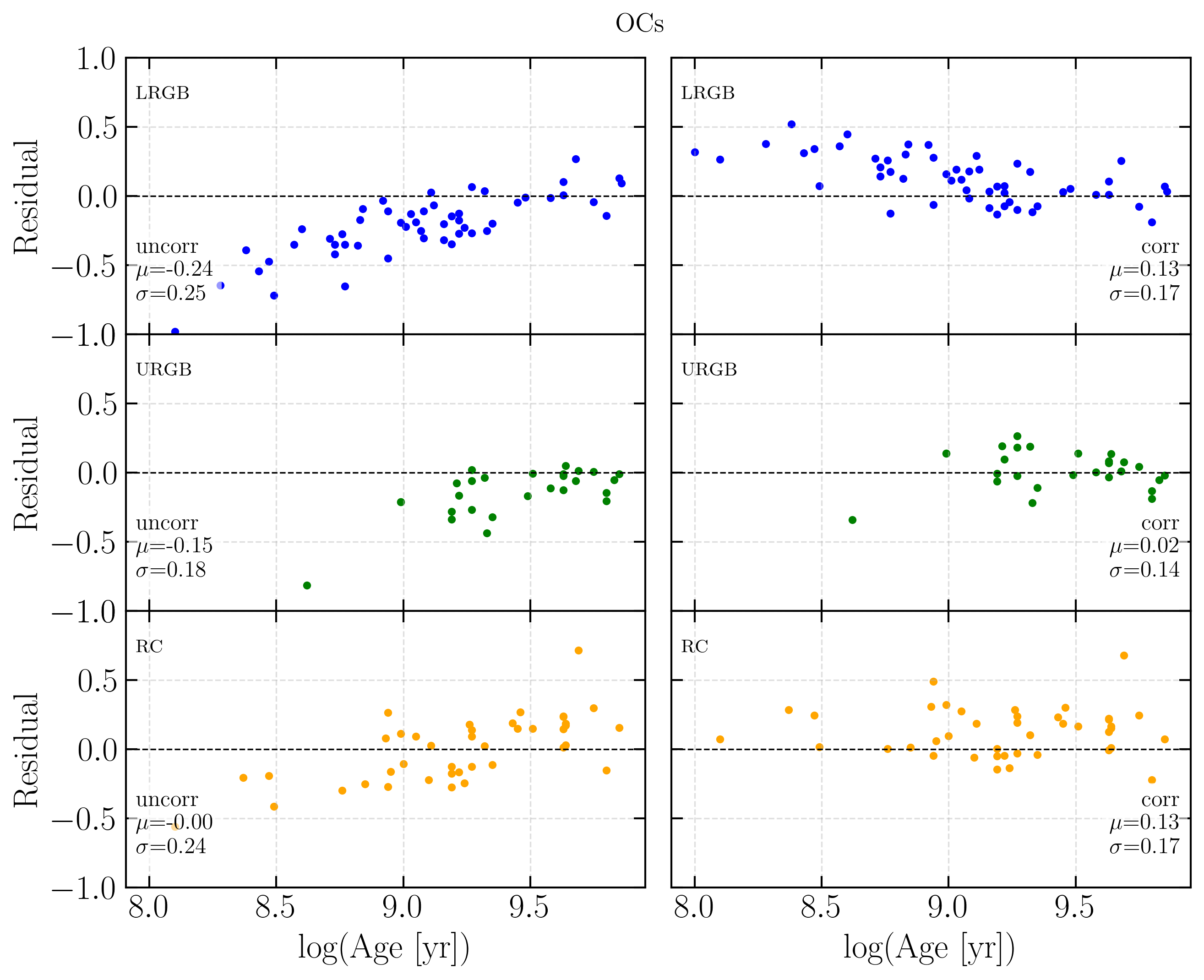}%{ocs_residuals_corr_vs_nocorr.png}
\caption{The residuals $\log(\mathrm{Age_{seismic}}) - \log(\mathrm{Age_{[C/N]}})$ vs $\log(\mathrm{Age_{seismic}})$ are shown for each open-cluster subsample (LRGB, URGB, RC).The left panels are the uncorrected residuals, the right panels are the corrected ones. \label{fig:residuals_ocs}}
\end{figure*}

We apply the\, \emph{Kepler}-based [C/N]-age-[Fe/H] calibrations to the independent TESS, K2, and open clusters (OCs) samples, considering the corresponding LRGB, URGB, and RC subsamples. For each dataset, we compare the [C/N]-based ages with the reference ages, adopting seismic ages for the field-star samples (TESS and K2) and isochrone-fitting ages for the OCs. The resulting residuals log(Age)$_{\rm [C/N]}$ $-$ log(Age)$_{\rm ref}$ are shown in Figs.~\ref{fig:residuals_tess}, \ref{fig:residuals_k2}, and \ref{fig:residuals_ocs}. In each figure, the left panels indicate the uncorrected [C/N]-based ages, while the right panels show the ages after applying the polynomial corrections derived from the \kepler sample.

For the K2 and TESS samples, the LRGB and URGB subsamples show residuals that are, on average, closer to zero with relatively small scatter ($\sigma \lesssim 0.2$~dex). Applying the correction derived for the \kepler sample further reduces the scatter and brings the mean closer to zero. However, the corrected model -- on average -- tends to underestimate the input ages in the validation sample. \\ %while it tends to overestimate the ages in the calibration sample.
For both data samples, the LRGB and URGB residuals deteriorate at old ages ($\log(\mathrm{Age/yr}) \gtrsim 10$) and young ages ($\log(\mathrm{Age/yr}) \lesssim 9.3$), as we already saw for the \kepler sample. In particular, for stars younger than $\sim2$~Gyr, the scatter increases by a factor of $\sim1.3$, which can significantly affect the overall mean and standard deviation. 
This residual deviation for young stars is most evident in the LRGB, since the URGB subsample contains relatively few stars in this young-age regime. 
As a result, the global statistics for URGB stars may appear comparable to, or even slightly better than, those of the LRGB sample, despite the fact that the calibration is less reliable at young ages. 
The comparatively good performance of the LRGB calibration (at least in the intermediate-age regime) is physically expected. These stars have not yet undergone the extra-mixing processes associated with the RGB-bump, which can substantially modify the surface [C/N] abundance in URGB phase. The relatively better agreement between seismic and [C/N]-based ages in this evolutionary phase therefore suggests that the [C/N]-age calibration is most robust for LRGB stars, especially at intermediate ages. Moreover, the seismic ages for URGB may be less accurate because of the difficulties associated with using \numax and \dnu at low frequencies, as is the case for stars on the upper red giant branch.
%By contrast, the URGB sample likely includes some contamination from stars near or beyond the onset of extra mixing, where larger scatter is expected. This leakage may partially mask the intrinsic differences between the LRGB and URGB regimes, leading to similar average residuals and dispersions.

%\mkn{see my suggestion in the figure caption for a few more tests with the residuals to see if they correlate with any variables you have in hand to check as well as comment on their magnitude with respect to the expectation from uncertainty alone.}
In contrast, the mean residuals for the RC subsamples are generally larger and the scatter increases modestly relative to the RGB subsamples. The discrepancy at young ages is, however, similar to that observed for the LRGB stars ($\log \rm (Age/yr) \lesssim 9.3$), while the discrepancy at old ages is bigger and more pronounced in the TESS sample. This is consistent with the expectation that additional evolutionary effects, such as mixing associated with core-helium burning, prior mass loss, or binary interactions, may affect both the surface [C/N] abundance and the seismic age estimates. The latter is expected because seismic masses (and hence ages) trace the present-day stellar mass, whereas [C/N]-based masses are more closely related to the mass at the time of the first dredge-up, before any subsequent mass loss or binary interaction \citep{roberts26b}.
%In particular, larger residuals for old RC stars may be driven by mass-loss effects, which are expected to be more significant for RC and may bias the inferred seismic masses and ages against the [C/N]-based ages. The systematically larger $\sigma$ values in the RC phase, in the old-age regime particularly, might therefore indicate a weaker correlation between [C/N] and age for these stars, but also that seismic ages are not fully reliable for comparison with [C/N]-based ages, due to effects such as mass loss or binary mass transfer. In fact, seismic masses (and hence ages) provide information on the present-day stellar mass, whereas [C/N]-based masses trace the stellar mass at the time of the FDU, before any possible mass loss or mass transfer \citep{roberts26b}.

The offset observed between the three different datasets, \kepler, K2 and TESS, could be related to their different age distributions and uncertainties, with TESS providing the least precise age estimates ($\sim 28\%$ against $\sim 20\%$ of \kepler and K2). As shown in Fig.~\ref{fig:hist}, the age distributions differ slightly among the three datasets and across the three evolutionary stages (LRGB, URGB and RC), with the K2 sample showing a somewhat older population. This difference may partly reflect the different Galactic regions probed by the K2 mission, which span a larger range in $R_{GC}$. Such differences in the underlying age distributions and age uncertainties may contribute to the observed differences in the residual trends in Figs.~\ref{fig:residuals},~\ref{fig:residuals_k2} and~\ref{fig:residuals_tess}, with the \kepler\, sample showing smaller residuals, as the relations were calibrated on this dataset. Nevertheless, the relations calibrated on the \kepler sample yield comparable results when applied to the K2 and TESS samples.

The OCs sample extends to younger ages than the \kepler, K2, and TESS samples, and correspondingly shows larger residuals at the youngest ages. In particular, for OCs with LRGB members, a clear trend is visible below $\log(\mathrm{Age/yr}) \simeq 9.3$, where the agreement between [C/N]-based and reference ages progressively worsens. This also corresponds to a regime that is not fully represented in the \kepler calibration sample for LRGB stars ($\log(\mathrm{Age/yr}) < 8.8$), even if we included OCs in the regression. %so the inferred [C/N]-based ages are effectively extrapolations. 
The same effect is less pronounced for the RC subsample, for which the \kepler calibration extends to younger ages (down to $\log(\mathrm{Age/yr}) \sim 8$). In addition to the limitations of the calibration itself, the discrepancies at young ages may also reflect genuine astrophysical effects, such as rotation or convective-core overshooting during the main-sequence phase, that are not fully captured by the present [C/N]-age relationship. These effects are discussed in Sec.~\ref{subsec:young}.\\
The corrected residuals show an opposite trend for the LRGB sample, driven by the larger discrepancy in the uncorrected values in the younger-age regime. For the URGB sample, the mean corrected residual is close to zero, while for the RC sample it remains positive, indicating an overestimation of the [C/N]-based ages with respect to the isochrone-based ones.

Overall, these results confirm that [C/N] can serve as a useful age proxy for giant stars, with the best performance obtained for LRGB stars in the intermediate-age regime. However, systematic effects associated with extra mixing may affect URGB and RC stars \citep[an effect that becomes even stronger at low metallicity; see][]{shetrone19}. In addition, the limitations of using \numax and \dnu at low frequencies, as is the case for stars on the upper red giant branch, can lead to less accurate asteroseismic age estimates for URGB stars, which may explain the large residuals. Mass loss and mass transfer, particularly for RC stars, may also contribute to the large residuals with respect to the seismic ages, as the latter trace the present-day stellar masses rather than the masses at the time of the first dredge-up, which are traced by [C/N]. However, investigating these effects is beyond the scope of this paper. They are discussed in more detail in \cite{roberts26b} \citep[see also][and papers therein for a detailed discussion of RGB mass loss]{brogaard24}. 
%{\bf As a test, we apply the most reliable relation derived for LRGB stars to the URGB and RC stars as well. We do not find any improvement in the agreement between the [C/N]-based and asteroseismic ages for the \kepler, K2, and TESS samples. Although different relations may be expected for more evolved stars because of mass loss and extra mixing, our approach allows us to verify that given the current asteroseismic age uncertainties, the LRGB relation may still be applicable to URGB and RC stars within the uncertainties.}
The method also becomes significantly less reliable for young populations, where extrapolation and additional stellar-physics effects likely dominate, and where complementary age diagnostics may therefore be required.

\section{Validity and Limitations of the [C/N]-Age relationship}
\label{sec:validityandlimits}

As shown in the previous sections, [C/N] serves as a more reliable age proxy for LRGB stars than for URGB or RC stars. Furthermore, the bias in age estimation is larger for young stars ($\lesssim 2$ Gyr) and old stars ($\gtrsim 10$ Gyr), compared to intermediate-age ones.
In terms of precision, ages derived from [C/N] for LRGB stars older than $\sim 2$ Gyr are precise to approximately 30\% (slightly larger for the oldest stars), decreasing to roughly 50\% for younger stars. For URGB and RC stars, the precision is generally lower: around 40\% for intermediate and old stars, and about 60\% for young RC stars. For URGB stars in the young regime, we lack sufficient data to draw conclusions.

In the following subsections, we discuss the limitations of the relationship and the regimes where it breaks down: the metal-poor regime, where extra-mixing effects are significant, the young regime and the oldest regime.

%\SB{Interesting idea for the metal-poor regime: What about the evolutionary carbon corrections from Vini Placco? \url{https://ui.adsabs.harvard.edu/abs/2014ApJ...797...21P}. I have not seen these discussed by Masserone2015 or Martig2015, although I believe they are somewhat important here to acknowledge - especially for the metal-poor regime. I am not sure if they are basically correcting for the effect that you try to use to measure age. But in any case, I think the paper should be somehow referenced when talking about the metal-poor regime. Happy to chat about this.}

%\mkn{need params regress over ranges shown in data section - maybe in histograms so Teff, logg, [Fe/H]}

\label{sec:limits}
\subsection{Metal poor regime and extra mixing}
\begin{figure*}
\centering
\includegraphics[scale=0.5]{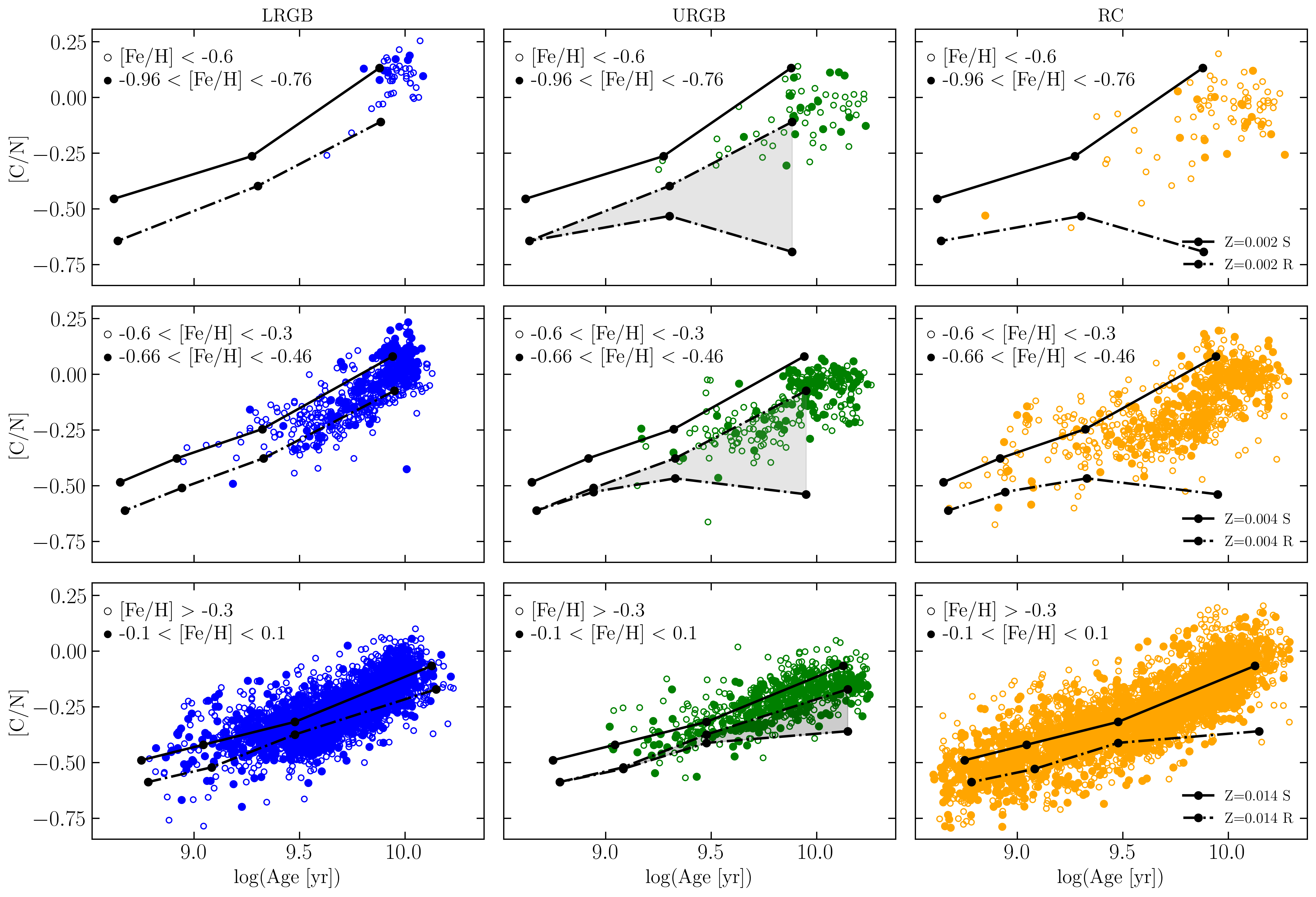}
\caption{Comparison between the predicted and observed trends of [C/N] as a function of the logarithm of stellar age. The data consist of \kepler stars (LRGB, URGB, RC) divided into three bins of [Fe/H]. Filled circles represent stars within $\pm 0.1$~dex of the corresponding model [Fe/H]. Theoretical models from \citet{lagarde12} are shown for three metallicity Z values. The curves represent models with standard prescription (S, solid line) and rotation-induced mixing and thermohaline instability (R, dot-dashed line). Black points on the model curves indicate the values corresponding to stellar masses of $1, 1.5, 2, 2.5 M_{\odot}$. %\SB{Sorry for asking tons of questions, but is there a reason to specifically cut a) only and b) specifically at {[Fe/H] = -0.4}? Given the significant amount of stars, I was wondering why not also dissect the sample into more abundance bins? Just because Largarde2012 does cover 0.004Z and 0.014Z should not hold you back from more bins - with some of them not having the theoretical models overlaid :)} 
\label{fig:extramixing}}
\end{figure*}

Following \citet{shetrone19}, giant stars beyond the RGB-bump are expected to exhibit signatures of extra mixing, particularly at low metallicity ($\rm [Fe/H]<-0.4$ dex). To investigate this effect, we compare the full \kepler sample, divided into three metallicity bins ($\rm [Fe/H] < -0.6, -0.6 < [Fe/H] < -0.3, [Fe/H] > -0.3$), across the three evolutionary stages considered in this work: LRGB, URGB, and RC (see Fig.~\ref{fig:extramixing}). We compare the observed surface abundances [C/N] vs log(Age/yr) with predictions from stellar evolution models.
The theoretical models adopted here are from \citet{lagarde12}, including both standard (S) and extra-mixing (R) prescriptions, which account for rotation-induced mixing and thermohaline instability. The grids cover initial masses between $1 M_{\odot} < M < 2.5 M_{\odot}$ and metallicities $Z = 0.002, 0.004, 0.014$ (corresponding to $\rm [Fe/H]=-0.86, -0.56 ~and~ 0$ dex, respectively). 
Since C and N abundances at birth vary with Galactic chemical evolution, while stellar models typically assume solar-scaled initial compositions, we apply the corrections of \citet{vincenzo21} to account for non-solar birth [C/N] ratios.

For each evolutionary phase, we extract representative model [C/N] values: (i) the post-FDU plateau for LRGB stars, where [C/N] becomes constant; (ii) the range between the RGB bump and the RGB tip for URGB stars (shown as a shaded area in Fig.~\ref{fig:extramixing}); and (iii) the value corresponding to the red clump phase. These three values are the same for standard models.
As shown in Fig.~\ref{fig:extramixing}, the $1M_{\odot}$ R model at the RGB tip and RC (point at the oldest age) exhibits a very strong extra-mixing signature compared to the other masses, which is not supported by the data. This shows that \citet{lagarde12}'s models over-predict extra mixing at old ages (lower masses), perhaps because of the numerical or physical treatment of thermohaline instability that strongly decreases the [C/N] content after the RGB bump.
% (see, e.g., Constantino et al. 2015, 2016; Constantino, Campbell, & Lattanzio 2017).}
%To mitigate this effect, we assume that the S-R difference varies linearly with mass and extrapolate this trend to $1M_{\odot}$, using a linear fit of the higher-mass points. The corrected value is then obtained by applying this extrapolated offset to the standard model prediction (red line in Fig.~\ref{fig:extramixing}).

Overall, LRGB stars show no significant signatures of extra mixing in any metallicity bin and are well reproduced by standard models (left panels of Fig.~\ref{fig:extramixing}). In contrast, URGB and RC stars exhibit clear evidence of extra mixing in the two sub-solar metallicity bins (top and middle panels), in agreement with theoretical expectations, while no significant effect is observed at solar metallicity, which defines the main sample of this work.
Moreover, in the most metal-poor bin, LRGB stars exhibit systematically higher mean [C/N] values compared to URGB and RC stars, suggesting that additional processes may affect the surface abundances in more evolved phases.

This analysis indicates that URGB and RC stars in the metal-poor regime are not reliable age tracers using [C/N], since post-RGB bump mixing processes alter their surface abundances and break the [C/N]-age-[Fe/H] relationship.
%\Giada{we need to say something about the difference in LRGB between S and R model. This means that there is an extra mixing before bump: rotation in MS?}

\subsection{Young regime ($\lesssim$ 2 Gyr)}
\label{subsec:young}
%Another limitation of this relationship concerns young stars. As seen from the residuals in Fig.~\ref{fig:residuals}, the relationship does not work properly for log(Age) < 9.3 (i.e., $\sim$ 2 Gyr). This is related to the different core structure of the star during its main-sequence phase. Indeed, young and massive stars have a convective core, and the nuclear reactions are dominated by the CNO cycle rather than the pp-chain. Since the efficiency of the CNO cycle depends strongly on temperature ($\propto T^{17}$, \textcolor{red}{[REF?]}), the reactions are concentrated in the central regions of the core. When the first dredge-up (FDU) occurs in these stars, it does not bring any processed material to the surface, and no chemical variations are observed. This explains why there is no correlation between age and [C/N] for youngest stars.
%\textcolor{red}{Overshooting? Rotation?}
Another limitation of the [C/N]-age-[Fe/H] relationship is observed at young ages, as highlighted by the residuals in Fig.~\ref{fig:residuals}, where the relationship progressively loses predictive power for $\log(\mathrm{Age/yr}) \lesssim 9.3$ (i.e. $\sim$2 Gyr). In this regime, stars are more massive and develop convective cores during their main-sequence evolution. %Their structure is dominated by the CNO cycle rather than the pp-chain, with nuclear reactions more strongly concentrated in the central regions due to the steep temperature dependence of the CNO cycle ($\propto T^{17}$). 
As a consequence, their surface abundances are more sensitive to complex internal transport processes, including convective-core overshooting and rotation-induced mixing, which depend on the distribution of initial rotational velocities.
In this mass regime, the sensitivity of surface [C/N] to stellar age is reduced, since the amount of C and N processed material available to be dredged up becomes less dependent on mass (e.g. \citealt{salaris15,roberts24}). 
At the same time, larger uncertainties in mixing processes near convective boundaries and in radiative regions (overshooting from cores, rotationally induced mixing) further contribute to the observed scatter in the relation at the young regime.
Moreover, stellar ages themselves become increasingly uncertain at higher masses, as different evolutionary models show a stronger dependence on assumptions such as overshooting and rotational mixing (see also \citealt{salaris15}). For these reasons, [C/N] is not expected to provide a tight age diagnostic for young, relatively massive stars, naturally explaining the loss of correlation at $\log(\mathrm{Age}) \lesssim 9.3$.
Although [C/N] is not a precise age diagnostic for stars younger than 2 Gyr, it can still reliably indicate whether a star belongs to the young regime.

\subsection{Old regime ($\gtrsim$ 10 Gyr)}
%For older stars, the [C/N]-age relation loses discriminating power because RGB stars span only a narrow range of low masses, so the expected variation in post-FDU [C/N] becomes too small compared to the observational and intrinsic scatter. Indeed, the relationship in this regime is nearly flat. However, it remains more reliable than in the young-age and metal-poor regime.
For older stars, the [C/N]-age-[Fe/H] relationship progressively loses discriminating power because RGB stars span only a narrow range of low masses, so that the expected variation in post-FDU [C/N] becomes comparable to, or smaller than, the combined observational and intrinsic scatter. In this regime, the relation is therefore nearly flat.
In addition, part of the observed dispersion may arise from variations in the birth [C/N] composition, which are expected to be more significant for old, metal-poor populations \citep[e.g.][]{kraft94,vincenzo21,roberts24,willett25}. These variations further increase the scatter and reduce the ability of [C/N] to act as a precise age diagnostic.
Although [C/N] is not a precise age diagnostic for stars older than 10 Gyr, it can still reliably indicate whether a star belongs to the old regime.

\section{Comparison with other dating methods}
\label{sec:comparisoncc}
In this section, we compare stellar ages derived from chemical clocks ([C/N], [Ce/Mg], [Zr/Ti]) with those obtained from asteroseismology and machine-learning techniques (i.e., astroNN and XGBoost) across different datasets. 
The analysed samples include: (i) 68 \textit{Kepler} stars from \citet{casali25} with high-resolution spectra from HARPS-N@TNG and FIES@NOT spectrographs and asteroseismic ages derived from individual mode frequencies using AIMS tool \citep[with a typical precision of $\sim$10\%,][]{aims}; (ii) the \textit{Kepler} stars used in this work with APOGEE abundances and asteroseismic ages inferred with PARAM (with a typical precision of $\sim$20\%); and (iii) the APOGEE dataset with ages estimated using the astroNN \citep{leung23} and (iv) XGBoost \citep{anders23} frameworks.
These datasets are limited to the \logg interval for giant stars from 2.4 to 3.5 dex (LRGB stars) and $\rm [Fe/H] > -1$ dex. The corresponding chemical ages are computed using the relationships reported in Table~\ref{tab:fit} for [C/N], and in Table~4 of \citet{casali25} for [Ce/Mg] and [Zr/Ti]. 

To estimate the accuracy and precision of the chemical ages, we calculate the mean and standard deviation of the relative errors of
the chemical ages compared to the input ages (asteroseismology or machine learning):
$$\rm \frac{Age_{comp} - Age_{chemical}}{Age_{comp}},$$
where $\rm Age_{chemical}$ are chemical ages from different chemical clocks, while $\rm Age_{comp}$ are ages from AIMS, PARAM, astroNN or XGBoost, respectively.
In this comparison, we do not show ages younger than 2 Gyr, as this is the regime in which [C/N]-based ages are most unreliable.
We interpret the mean of these relative errors as a measurement of the accuracy, while the standard deviation as a measurement of the precision of chemical ages. 

For the ages inferred from [C/N], we also show the corrected ages obtained using the correction derived in Fig.~\ref{fig:residuals}, where $\rm Age_{comp}$ is adopted as the reference age.

\begin{figure*}
\centering
\includegraphics[scale=0.4]{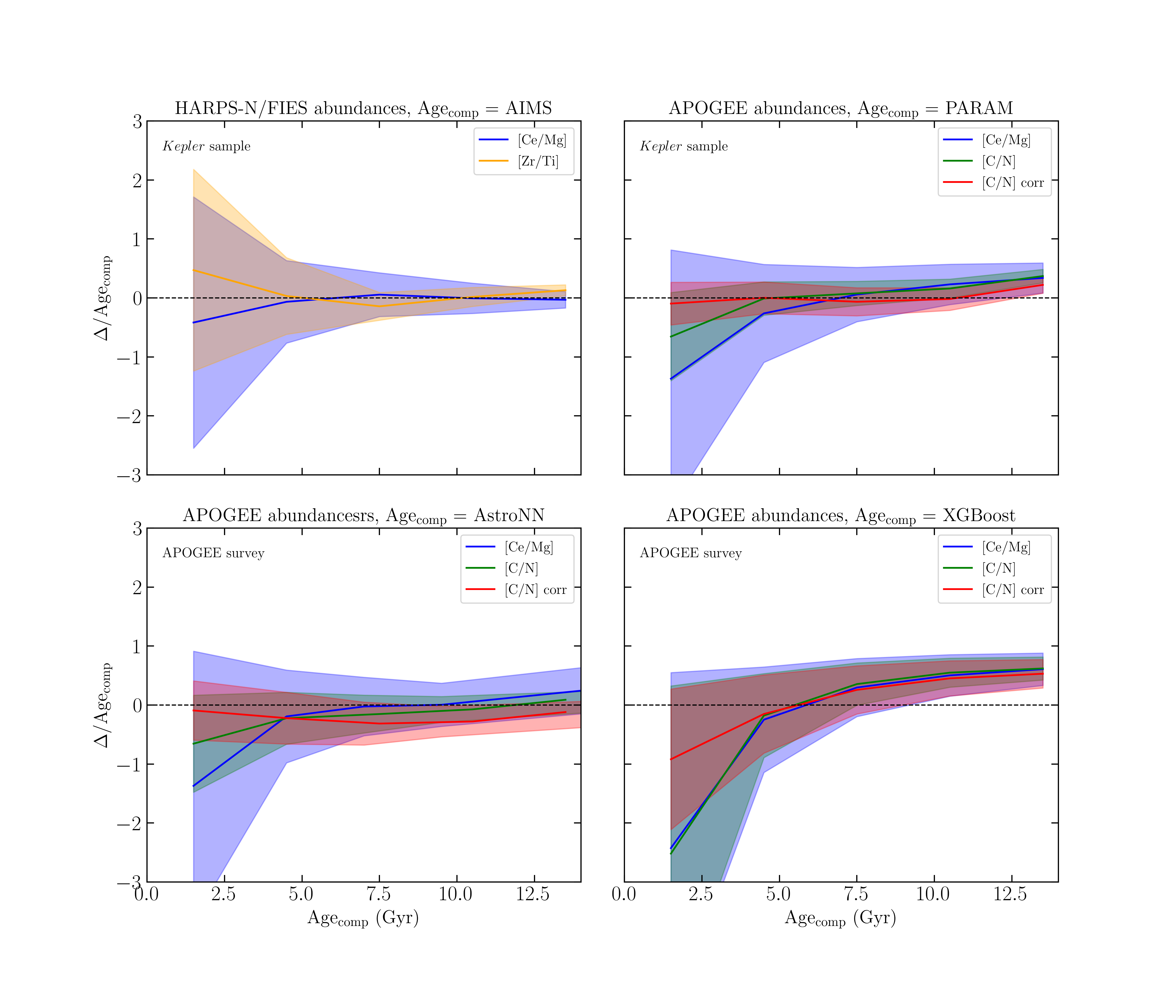}
\caption{The mean (solid line) and standard deviation (shaded area) of
$\Delta$/Age$_{\rm comp}$ in different age bins, with $\rm \Delta = Age_{comp} - Age_{chemical}$. The labels $\rm Age_{chemical}$ and Age$_{\rm comp}$ are explained in the legend and panel's title, respectively. \label{fig:comparison}}
\end{figure*}

From Fig.~\ref{fig:comparison}, we note that the sample with the highest spectroscopic precision \citep[top-left panel, 68 \kepler stars from][]{casali25} exhibits the best agreement between chemical and asteroseismic ages. This demonstrates the critical role of precise abundance measurements in achieving more accurate and precise chemical age estimates. However, for this high-resolution sample, C and N abundances are not available, preventing a direct comparison with the [C/N]-based relationships.

When comparing asteroseismic ages with those derived from [Ce/Mg] and [C/N] using the APOGEE abundances (top-right panel, \kepler sample from this work), the [C/N]-based ages show both higher precision and better agreement with input seismic ages with respect to [Ce/Mg]-based ages. This could reflect two factors: (i) C and N abundances are generally more reliable than Ce in APOGEE survey, considering that neutron-capture elements, such as Ce, are more difficult to measure in the infrared (the APOGEE spectral range) than in the optical, and (ii) the [C/N]-age-[Fe/H] relationship exhibits a tighter correlation than the corresponding [Ce/Mg]-age-[Fe/H] relationship. Moreover, chemical clocks related to Galactic evolution (e.g., [Ce/Mg]) depend not only on metallicity but also on the location within the Galaxy, adding another source of uncertainty \citep[see, e.g.,][]{casali20,casali23,viscasillas22,ratcliffe23,ratcliffe24,molero25}.
For both chemical age estimates, the accuracy decreases for ages younger than 4 Gyr and older than 10 Gyr. When the correction is applied, the [C/N]-based ages become more accurate (see the red solid line).

Comparisons between chemical ages and those from machine-learning methods (lower panels) show a systematically poorer agreement with respect to those with asteroseismic ages (upper panels). This is likely due to the lower intrinsic precision of the machine-learning age estimates with respect to the asteroseismic ones. Even in these cases, the [C/N]-based ages remain more precise and accurate than those from [Ce/Mg]. This is probably also due to the lower precision of APOGEE Ce abundances compared to C and N. 
The comparison with XGBoost in particular reveals a significant discrepancy for both chemical clocks. Moreover, the [$\alpha$/Fe]-[Fe/H] diagram colour-coded by XGBoost ages does not show any clear dichotomy in stellar ages between the high- and low-$\alpha$ sequences, suggesting potential issues with the XGBoost age determinations.

\begin{figure}
\centering
\includegraphics[scale=0.45]{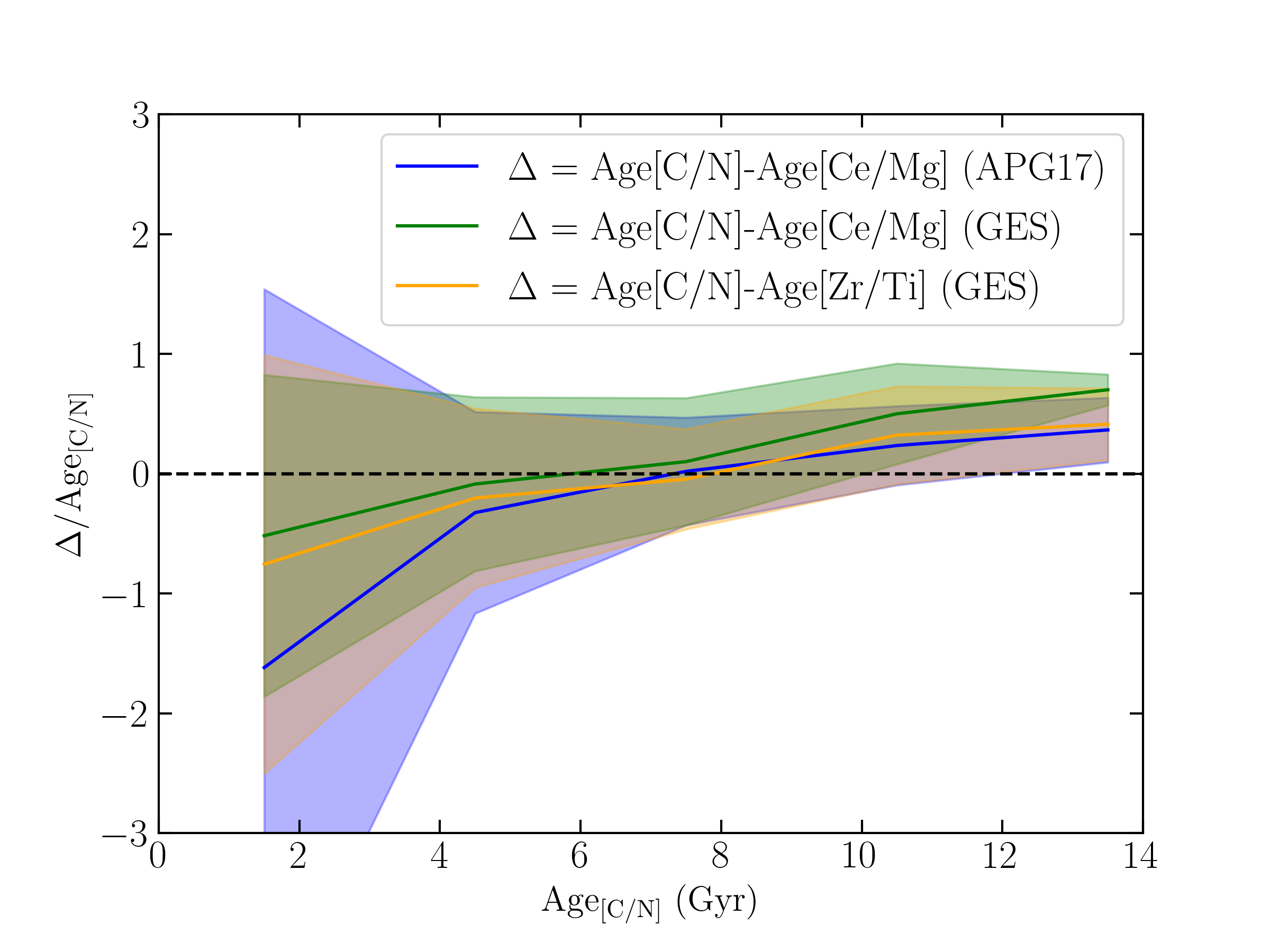}
\caption{The mean (solid line) and standard deviation (shaded area) of
$\Delta$/Age$_{\rm [C/N]}$ in different chemical age bins, where $\rm \Delta$ is explained in the legend. \label{fig:comparison2}}
\end{figure}

Figure \ref{fig:comparison2} shows the comparison between stellar ages derived from different chemical clocks across two spectroscopic surveys: APOGEE and \emph{Gaia}-ESO. For the \emph{Gaia}-ESO sample, we apply abundance offsets for the elements in common with APOGEE (i.e., C, N, Mg, Ti, Ce and [Fe/H]) to account for systematic differences between the two surveys. The comparisons involve the pairs [C/N]-[Ce/Mg] and [C/N]-[Zr/Ti].
The plot highlights that the agreement between ages inferred from [Ce/Mg] and [C/N] is poorer for APOGEE than for the \emph{Gaia}-ESO survey. In APOGEE, both the accuracy and the precision are reduced (particularly at young ages), suggesting lower-quality abundances of C, N, and Ce compared to the \emph{Gaia}-ESO survey.
For the \emph{Gaia}-ESO sample, both comparisons ([C/N]-[Ce/Mg] and [C/N]-[Zr/Ti]) show a degradation of agreement at older ages. This is mainly driven by the limited number of \emph{Gaia}-ESO stars in this regime with reliable measurements of all required abundances and surface gravities for LRGB stars.
Within the \emph{Gaia}-ESO data, the [C/N]-[Zr/Ti] comparison performs slightly better than [C/N]-[Ce/Mg], suggesting that [Zr/Ti] may trace evolutionary timescales more consistently.

\section{Application to field stars}
\label{sec:application}
\begin{figure}
\centering
\includegraphics[scale=0.45]{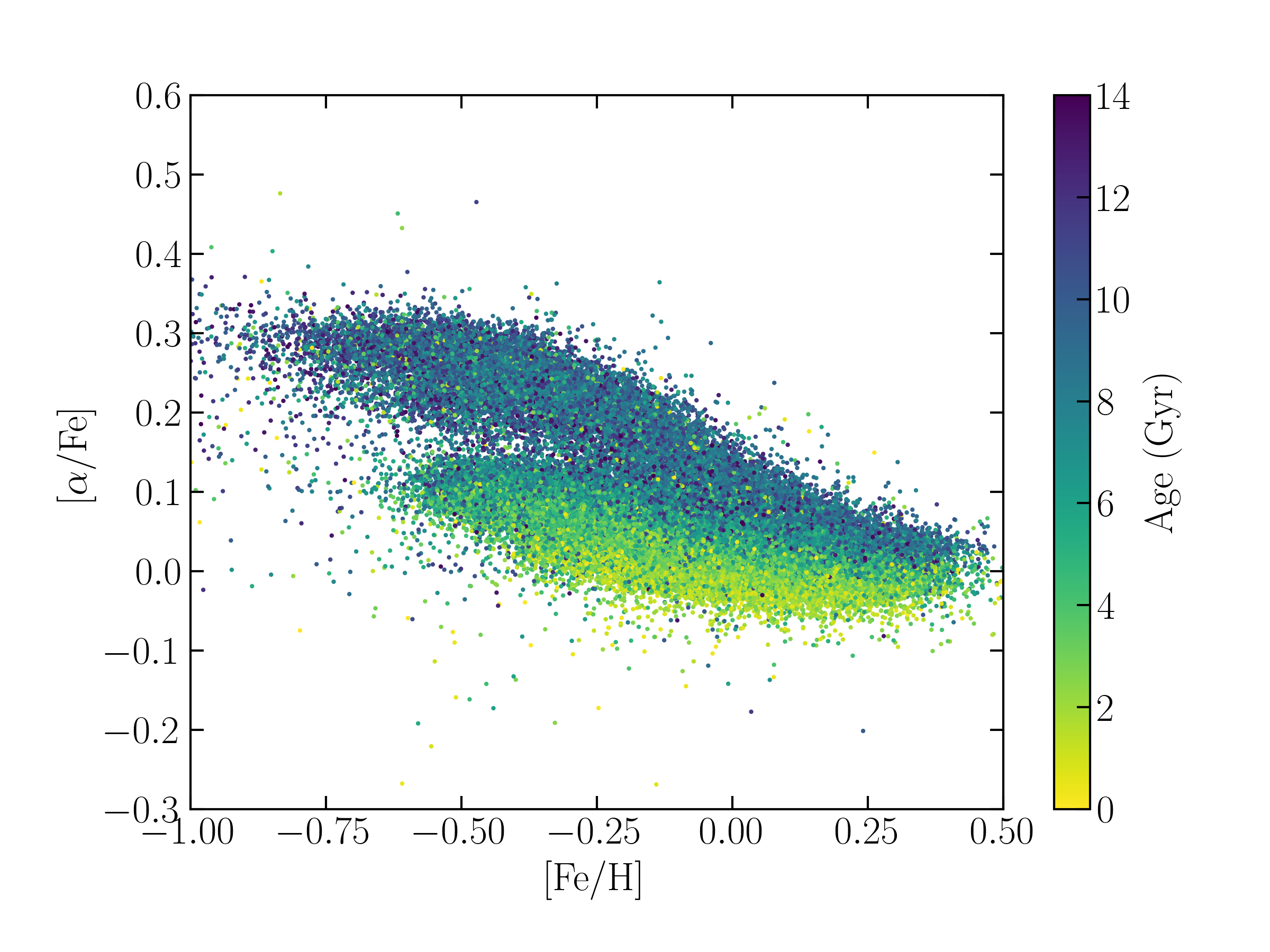}
\caption{[$\alpha$/Fe] vs [Fe/H] of lower-RGB stars ($\rm 2.4 < \log~g < 3.5, [Fe/H] > -1$) in the APOGEE survey. Stars are colour-coded by ages based on [C/N]. \label{fig:afe_feh}}
\end{figure}

\begin{figure}
\centering
\includegraphics[scale=0.45]{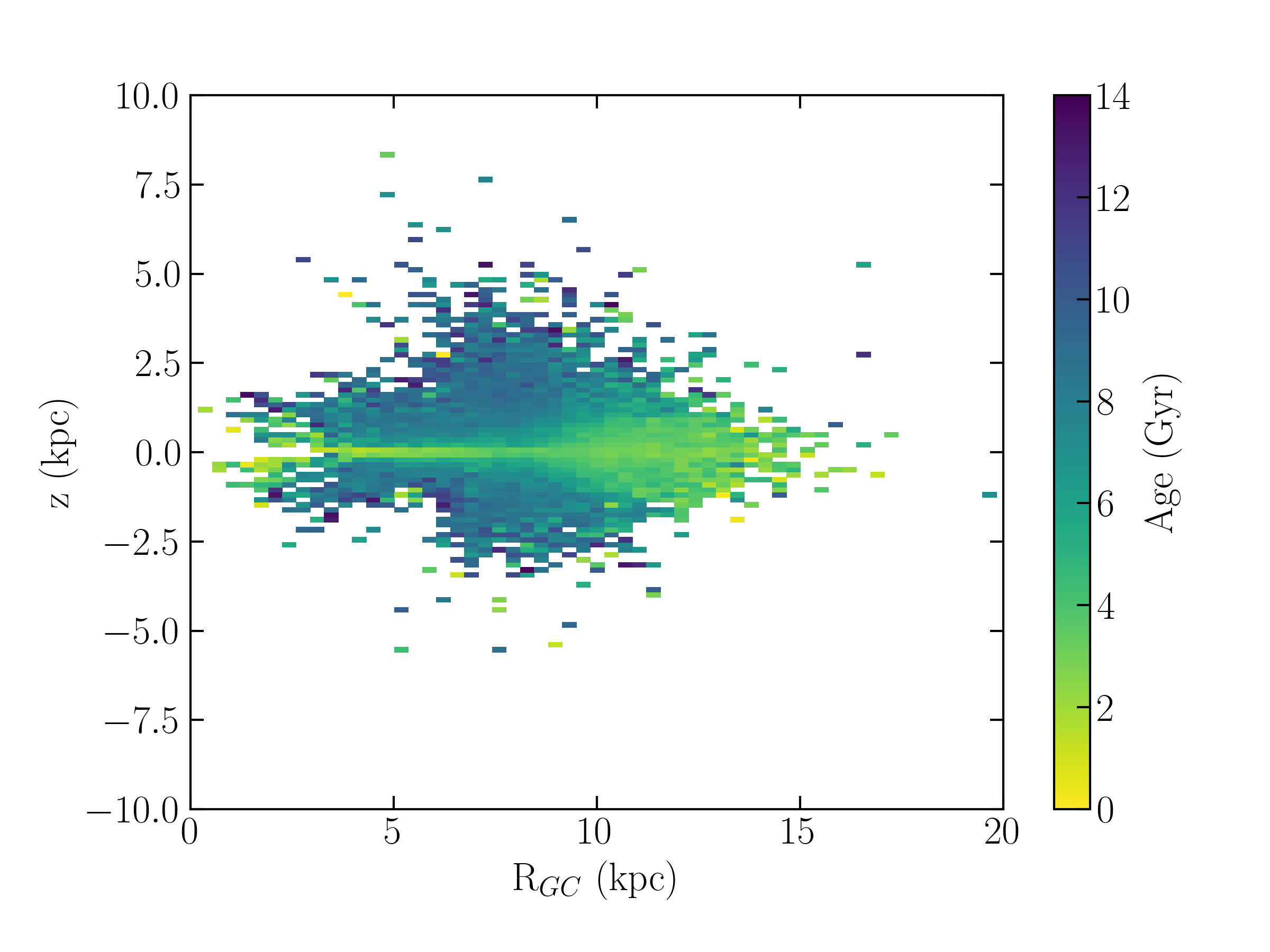}
\caption{$z$ vs $R_{GC}$ of lower-RGB stars ($\rm 2.4 < \log~g < 3.5, [Fe/H] > -1$) in the APOGEE survey. Stars are colour-coded by ages based on [C/N]. \label{fig:flaring}}
\end{figure}

We finally apply the relationship found in this work to all LRGB stars present in the APOGEE survey, selecting stars in the \logg interval $2.4 < \log~g < 3.5$~dex. %considering the limit in \logg discussed in the previous sections. 
As we can see from Fig.~\ref{fig:afe_feh} and Fig.~\ref{fig:flaring}, chemical ages from [C/N] describe well some key features of the Galactic archaeology.

Figure~\ref{fig:afe_feh} shows the [\alfa/Fe] vs [Fe/H] plane colour-coded by chemical ages from [C/N] ratios. The dichotomy between high-\alfa~and low-\alfa~sequences is quite evident: high-\alfa~stars are systematically older with a mean age around 11 Gyr, whereas low-\alfa stars show a gradient in age going to higher \alfa~content.

Figure~\ref{fig:flaring} shows the distribution of stars in the \rgc vs $z$ plane, where \rgc represents the Galactocentric radius and $z$ represents the height above the Galactic plane. The distribution is colour-coded by the chemical ages derived from the [C/N] ratio. The map clearly displays the Milky Way’s disc flaring: stars located close to the Galactic plane are generally young, while the vertical extent of the stellar distribution increases at larger Galactocentric distances. Conversely, older stars tend to reside at smaller \rgc and at greater distances from the plane than their younger counterparts.
The concentration of the youngest stars near the plane points to ongoing star formation in the gas-rich regions of the disc. Their presence at progressively larger radii supports an inside-out formation scenario for the Milky Way \citep[e.g.,][]{matteucci89,chiappini01}, in which star formation began in the central regions and gradually propagated outward. This pattern aligns with the notion that the Galactic disc has expanded over time, leaving older populations concentrated toward the inner Galaxy and younger stars forming in the outer regions. Such a distribution provides valuable insight into the coupling between stellar ages and the structural evolution of the Milky Way.

These figures demonstrate how [C/N]-based ages are reasonably reliable for Galactic archaeology studies, enabling us to estimate ages for large samples of stars. This approach provides an efficient way to derive stellar ages from spectroscopic data alone, making it particularly valuable for large-scale surveys where other methods (e.g., asteroseismology) are not available for comparable numbers of stars.

\section{Summary and conclusions}
\label{sec:conclusions}
In this work, we calibrate and test the [C/N] abundance ratio as a chemical clock for giant stars, using asteroseismic ages from the \kepler mission combined with APOGEE DR17 abundances. 

We derive multivariate quadratic and linear relationships between $\log(\rm Age/yr)$, [C/N], and [Fe/H] for lower RGB (LRGB), upper RGB (URGB), and red clump (RC) stars. The quadratic relations generally outperform linear fits, as quantified by $R^2_{\rm adj}$, BIC, and AIC values. The calibration is most robust for LRGB stars, for which extra mixing has not yet significantly altered the surface [C/N] abundances.

We apply the \emph{Kepler}-based calibration to independent K2, TESS, and open clusters samples to validate our relationships. It confirms that the [C/N]-age-[Fe/H] relationship provides reliable age estimates, particularly for LRGB stars. For URGB and RC stars, residuals with respect to asteroseismic ages are larger and these stars exhibit weaker correlations between [C/N] and age.

The [C/N]-age-[Fe/H] relationship is less accurate for young stars ($\lesssim 2$ Gyr), where complex internal transport processes in the main sequence phase invalidate [C/N] as a robust age diagnostic, and for metal-poor URGB and RC stars ([Fe/H] < –0.4), where extra-mixing effects introduce systematic biases. 
For older stars ($\gtrsim 10$ Gyr), the [C/N]-age-[Fe/H] relationship progressively loses discriminating power, as RGB stars occupy a narrow range of low masses and are additionally affected by variations in their initial [C/N] composition.
However, [C/N]-based ages still enable a robust relative classification, with stars younger than $\sim 2$ Gyr reliably identified as young, and those older than $\sim 10$ Gyr safely classified as old.
Age uncertainties derived from [C/N] show a typical precision of $\sim 30\%$ for LRGB stars in the intermediate-age regime. This slightly degrades at older ages and decreases to $\sim 50-60\%$ in the young regime.

%Differences at the youngest and oldest ends highlight the importance of accounting for evolutionary stage, metallicity, and extra mixing when estimating ages from [C/N].

We compare chemical ages derived from [C/N] and [Ce/Mg] and find that [C/N] provides more reliable ages in the regime where it is valid, exhibiting smaller residuals and tighter correlations with asteroseismic ages. This is likely due to the less precise Ce abundances in the APOGEE survey, as well as the fact that [Ce/Mg] is influenced by Galactic chemical evolution and the enrichment history of neutron-capture elements, which can vary significantly with location and metallicity, introducing additional scatter. Therefore, [C/N] is physically a more robust stellar-age indicator for red giants.

The [C/N] chemical clock provides a valuable age proxy for red giant stars in large spectroscopic surveys, enabling the study of Galactic evolution over broad spatial scales. The method is especially effective for LRGB stars with [Fe/H] > –1, but caution is required for URGB, RC, young/old, or metal-poor stars.

Overall, our results demonstrate that [C/N], combined with metallicity information, offers a powerful empirical tool for deriving stellar ages in the Milky Way, complementing asteroseismic and isochrone-based techniques, and enabling robust age estimates for large field-star populations beyond the reach of current asteroseismic samples.

\section*{Acknowledgements}
GC, AM acknowledge
support from the European Research Council Consolidator Grant funding scheme (project ASTEROCHRONOMETRY, G.A. n. 772293, \url{http://www.asterochronometry.eu}.

\section*{Data availability}
The data sets used and analysed for this study are taken from \citet{willett25}, APOGEE DR17 \citep{apogeedr17}, Gaia-ESO \citep{randich22}, \citet{spoo22}, \citet{grazina25}, \citet{casali25}, \citet{leung23} and \citet{anders17}. The rest of the
relevant data sets are available from the corresponding author on
reasonable request.


\bibliographystyle{mnras}
\bibliography{Bibliography} % if your bibtex file is called example.bib

%\begin{thebibliography}{99}
%\end{thebibliography}

%%%%%%%%%%%%%%%%%%%%%%%%%%%%%%%%%%%%%%%%%%%%%%%%%%

%%%%%%%%%%%%%%%%% APPENDICES %%%%%%%%%%%%%%%%%%%%%

\appendix

%%%%%%%%%%%%%%%%%%%%%%%%%%%%%%%%%%%%%%%%%%%%%%%%%%

% Don't change these lines
%\bsp	% typesetting comment
\label{lastpage}
\end{document}